\documentclass[11pt,a4paper,twocolumn]{article}

\usepackage{graphicx}
\usepackage{epsfig}
\usepackage{natbib}
\usepackage{psfrag}
\usepackage{amsmath}
\usepackage{array}
\usepackage{lscape}
\usepackage{authblk}

\usepackage{color}
\definecolor{darkgreen}{rgb}{0.1, 0.5, 0.1}

\newcommand{\subtext}[1]{{\mbox{\scriptsize{#1}}}}

\newcommand{\refp}[1]{(\ref{#1})}

\newcommand{\B}{\mbox{$\mathbf B$}}
\newcommand{\U}{\mbox{$\mathbf U$}}

\newcommand{\be}{\begin{equation}}
\newcommand{\ee}{\end{equation}}
\newcommand{\eep}{\;\;.\end{equation}}
\newcommand{\eec}{\;\;,\end{equation}}
\newcommand{\bea}{\begin{eqnarray}}
\newcommand{\eea}{\end{eqnarray}}
\newcommand{\bel}[1]{\begin{equation}\label{#1}}
\newcommand{\beal}[1]{\begin{eqnarray}\label{#1}}
\newcommand{\tp}[1]{\mbox{$\times10^{#1}$}}

\newcommand{\uvr}{\hat{\mathbf r}}

\newcommand{\uvz}{\hat{\mathbf z}}

\newcommand{\Rol}{\mbox{Ro$_\ell$}}
\newcommand{\E}{\mbox{E}}

\newcommand{\Pra}{\mbox{Pr}}
\newcommand{\Pm}{\mbox{Pm}}

\newcommand{\Ra}{\mbox{Ra}}

\newcommand{\rhob}{\overline{\rho}}

\newcommand{\Figref}[1]{Fig.~(\ref{#1})}
\newcommand{\figref}[1]{fig.~(\ref{#1})}

\newcommand{\eqnref}[1]{eqn.~(\ref{#1})}
\newcommand{\Secref}[1]{Section \ref{#1}}
\newcommand{\secref}[1]{section \ref{#1}}

\newcommand{\secrefp}[2][]{(#1 section \ref{#2})}

\newcommand{\tabref}[1]{table {\ref{#1}}}
\newcommand{\Tabref}[1]{Table {\ref{#1}}}

\begin{document}


\title{Mercury's magnetic field in the MESSENGER era}

\author[1]{J. Wicht}
\author[2]{D. Heyner}
\date{\today}


\affil[1]{Max-Planck Institut f\"ur  Sonnensystemforschung,
                  G\"ottingen, Germany,
                  wicht@mps.mpg.de}
\affil[2]{Institut f\"ur Geophysik und extraterrestrische Physik, TU Braunschweig,
                  Braunschweig, Germany}


\twocolumn[\begin{@twocolumnfalse}
\maketitle

\begin{abstract}

MESSENGER magnetometer data show that Mercury's magnetic field is
not only exceptionally weak but also has a unique geometry.
The internal field resembles an axial dipole that is offset to the North
by 20\% of the planetary radius.
This implies that the axial quadrupol is particularly strong while the
dipole tilt is likely below $0.8^\circ$.
The close proximity to the sun in combination with the weak internal
field results in a very small and highly dynamic Hermean magnetosphere.
We review the current understanding of Mercury's internal and
external magnetic field and discuss possible explanations.
Classical convection driven core dynamos have a hard time to reproduce
the observations.
Strong quadrupol contributions can be promoted
by different measures, but they always go along with a large dipole
tilt and generally rather small scale fields.
A stably stratified outer core region seems required to
explain not only the particular geometry but also the weakness of
the Hermean magnetic field. New interior models suggest that
Mercury's core likely hosts an iron snow zone underneath the
core-mantle boundary. The positive radial sulfur gradient likely
to develop in such a zone would indeed promote stable stratification.
However, even dynamo models that include the stable layer show
Mercury-like magnetic fields only for a fraction of the total simulation time.
Large scale variations in the core-mantle boundary heat flux promise
to yield more persistent results but are not compatible with the
current understanding of Mercury's lower mantle.

\end{abstract}
\end{@twocolumnfalse}]



\section{Introduction}
\label{Intro}

In 1974 the three flybys of the Mariner 10 spacecraft revealed that
Mercury has a global magnetic field.
This was a surprise for many scientists since an internal dynamo process
was deemed unlikely because of the planet's relative small size
and its old inactive surface \citep{Solomon1976}.
Either the iron core would have already solidified completely
or the heat flux through the core-mantle boundary (CMB) would be too small
to support dynamo action.
The Mariner 10 measurements also indicated that Mercury's magnetic field
is special \citep{ness_1974}. Being $100$ times smaller than the geomagnetic
field, it seems too weak to be supported by an Earth-like core dynamo.
And though the data were scarce, they nevertheless allowed to constrain that the internal
field is generally large scale and dominated by a dipole but possibly also
a sizable quadrupole contribution.
Both the Hermean field amplitude and its geometry are unique in our solar system.

Mercury is the closest planet to the Sun and therefore subject to a
particular strong and dynamic solar wind. Since Mercury's magnetic field is
so weak, the solar wind plasma can come extremely close to the planet and
may even reach the surface. Mariner 10 data showed that Mercury's magnetosphere
is not only much smaller than its terrestrial counterpart but also much
more dynamic. Adapted models originally developed for Earth failed
to adequately describe the Hermean magnetosphere which therefore remained
little understood in the Mariner 10 era \citep{Slavin2007}.

Knowing a planet's internal structure is crucial for understanding the
core dynamo process. Mercury's large mean density
pointed towards an extraordinary huge iron core and a relatively thin
silicate mantle covering only about the outer 25\% in radius.
Since little more data were available in the Mariner era,
the planet's interior properties and dynamics remained poorly constrained.

Solving the enigmas about Mercury's magnetic field
and interior where major incentives for NASA's MESSENGER mission \citep{Solomon2007}.
After launch in August 2004 and a first Mercury flyby in January 2008,
the spacecraft went into orbit around the planet in March 2011.
At the date of writing, more than 2800 orbits have been completed.
MESSENGER's orbit is highly eccentric with a periapsis between $200$ to
$600\,$km at $60$ to $70^\circ$northern latitude and an apoapsis of about $15,000\,$km altitude.
This has the advantage that the spacecraft passes through the magnetosphere
on each orbit but complicates the extraction of the internal field component
because of a strong covariance of equatorially symmetric and
anti-symmetric contributions \citep{Anderson2012,Johnson2012}.
The tradeoff between the dipole and quadrupole field harmonics, that was already
a problem with Mariner 10 data, therefore remains an issue in the MESSENGER era.
The situation is further complicated by the fact that the
classical separation of external and internal field contributions
developed by Gauss \citep{Olsen2010} does not directly apply at Mercury.
It assumes that the measurements are taken in a
source free region with negligible electric currents, an assumption not
necessarily fulfilled in such a small and dynamic magnetosphere.

In order to nevertheless extract information on the internal magnetic field, the
MESSENGER team analysed the location of the magnetic equator where $B_\rho$,
the magnetic field component perpendicular to the planet's rotation axis,
passes through zero \citep{Anderson2011,Anderson2012}.
Since the internal field changes on a much slower time scale than
the magnetosphere, the time-averaged location should basically
not be affected by the magnetospheric dynamics.
The analysis not only confirmed that the Hermean field is exceptionally
weak with an axial dipole of only $190\,$nT but also suggested that the internal
field is best described by an axial dipole that is offset by $480\,$km
to the north of the planet's equator \citep{Anderson2012}.
This configuration, that we will refer to as the MESSENGER offset dipole
model (MODM) in the following, requires a strong axial quadrupole
and a very low dipole tilt, a combination that is unique in our solar system.

This article tries to summarize the new understanding of Mercury's
magnetic field in the MESSENGER era at the date of writing.
MESSENGER is still orbiting it's target and continues to
deliver outstanding data that will further improve our knowledge
of this unique planet.
\Secref{External} briefly reviews the current knowledge of
Mercury's magnetosphere. \Secref{Internal} describes recent models
for the planet's interior, focussing in particular on the possible
core dynamics. The magnetic equator analysis and the offset dipole model
MODM are then discussed in \secref{Internal}.
Explaining the weakness of Mercury's magnetic field already
challenged classical dynamo theory and the peculiar field geometry
further raises the bar. \Secref{Dynamo} reanalysis several dynamo model
candidates in the light of the new MESSENGER data.
Some concluding remarks in \secref{Conclusion} close the paper.

\section{Mercury's internal\\ structure}
\label{Interior}

MESSENGER observations of Mercury's gravity field \citep{Smith2012} and
Earth-based observations of the planet's spin state \citep{Margot2012}
provide valuable information on the interior structure.
That fact that  Mercury is in a special rotational state (Cassini state 1)
allows to deduce the polar moment of inertia $C$ from the
degree two gravity moments and the planet's obliquity, the
tilt of the spin axis to the orbital normal \citep{Peale1969}.
The moment of inertia factor $C/(M R_M^2)$, where $M$ is
the planet's total mass and $R_M$ its mean radius, constrains the
interior mass distribution. The
factor is $0.4$ for uniform density and decreases when the mass
is increasingly concentrated towards the center.
The Hermean value of $C/(M R_M^2) = 0.346\pm 0.014$ \citep{Margot2012} indicates a
significant degree of differentiation.

The observation of the planet's $88\,$day libration amplitude $g_{88}$,
a periodic spin variation in
response to the solar gravitational torques on the asymmetrically shaped
planet, allows to also deduce the moment of inertia of the rigid outer part
$C_m$. If the iron core is at least partially liquid, $C_m$ is the moment
of the silicate shell and thus smaller than $C$.
The Herman value of $C_m/C = 0.431\pm 0.025$ \citep{Margot2012}
confirms that the core remains at least partially liquid.

In addition to $M$ and $R_M$ the ratios $C/(M R_M^2)$ and $C_m/C$
provide the main constraints for models of Mercury's interior \citep{Smith2012,Hauck2013}.
Note that \citet{Rivoldini2013} follow at somewhat different approach,
taking into account the possible coupling between the core and the
silicate shell. The coupling has the effect that $C_m$ cannot be determined
independently of the interior model and \citet{Rivoldini2013} therefore
directly use $g_{88}$ rather than $C_m$ as a constraint.
The updated interior modelling indicates that the core radius is relatively well constrained
at $2020\pm30\,$km \citep{Hauck2013} or $2004\pm\,39$km \citep{Rivoldini2013}.
This leaves only the outer $16$ to $19$\% of the mean planetary radius
$R_M=2440\,$km to the mantle.

\citet{Hauck2013} find a mean mantle density (including the crust)
of $3380\pm200\,$kg/m$^3$.
Measurements of MESSENGER's X-Ray Spectrometer (XRS) show that the
volcanic surface rocks have a low content of iron and other
heavier elements \citep{Nittler2011}. \citet{Smith2012} and
\citet{Hauck2013} therefore speculate that a solid FeS outer core layer
may be required to explain the mean mantle density.
\citet{Rivoldini2013}, however, argue that the mantle density is not
particularly well contrained. Compositions compatible with XRS measurements
are well within the allowed solutions and a denser lower mantle layer
is not required by the data.

Naturally, information about the core is of particular interest for the
planetary dynamo. There is a rough consensus on the mean core
density with \citet{Hauck2013} and \citet{Rivoldini2013}
suggesting $6980\pm280\,$km/m$^3$
and $7233\pm267\,$km/m$^3$, respectively. However, the core
composition and the radius of a potential inner core are not well
constrained. Admissable interior models cover all inner core radii
from zero to very large values with an aspect ratio of about
$a=r_i/r_o=0.9$ \citep{Rivoldini2013} where $r_i$ and $r_o$ are the inner and
outer core radii, respectively.

An additional constraint on the inner core size relies on the observations
of so-called lobate scarps on the planet's surface which
are likely caused by global contraction. MESSENGER data based on $21$\% of the
surface suggested a contraction between $1$ and $3\,$km \citep{DiAchille2012}.
This sets severe bounds on the amount of solid iron in Mercury's core because
of the density decrease associated with the phase transition of the liquide
core alloy.
Several thermal evolution models therefore favour a completely liquide core
or only a very small inner core \citep{Grott2011,Tosi2014}. Recent more comprehensive
MESSENGER observations, however, allow for a contraction of up to $7\,$km.
This somewhat releases the contraints \citep{Solomon2013}
though very large inner cores may still be unlikely.

Sulfur has been found in many iron-nickel meteorites and is therefore
a prime candidate for the light constituent in Mercury's core.
\citet{Rivoldini2013} consider iron-sulfur core alloys
and find a likely
bulk sulfur concentration of $4.5\pm 1.8\,$wt\%.
Since this composition lies on the iron rich side of the eutectic,
iron crystalizes out of the liquid when the temperature drops
below the melting point. Where this happens first depends on the
form of the melting curve and the adiabat describing core conditions.

Since Mercury's mantle is so thin it has likely cooled to a point
where mantle convection is very sluggish or may have
stopped altogether \citep{Grott2011,Michel2013,Tosi2014}.
The heat flux through the core-mantle boundary is thus likely subadiabatic
and therefore too low to support a core dynamo driven by thermal convection
alone. The required additional driving power may then either be provided by a growing
inner core or by an iron snow zone. The solid inner core starts
to grow as soon as  the adiabat crosses the melting curve in the planetary center.
Since the solid iron phase can incorporate only a relatively
small sulfur fraction, most of the sulfur is expelled at the inner core front
and drives compositional convection.
The latent heat released upon iron solidification provides additional thermal
driving power. Contrary to the situation for Earth, freezing could also start
at the core-mantle boundary (CMB) because of the lower pressures in Mercury's core.
The iron crystals would then precipitate or snow into the center and remelt when
encountering temperatures above the melting point at a depth $r_m$.
This process leaves a sulfur enriched lighter residuum in the
layer $r>r_m$. As the planet cools, $r_m$ decreases and a stabilizing
sulfur gradient is established that follows the liquidus curve and
covers the whole snow zone $r>r_m$ \citep{Hauck2006}.
Since the heat flux through the CMB is likely subadiabatic today, thermal effects
will also suppress rather than promote convection in the outer part of
Mercury's core. A stably stratified layer underneath the
planet's core mantle boundary and probably extending over
the whole iron snow region therefore seems likely.
The liquid iron entering the layer below $r_m$ serves as a compositional
buoyancy source. The latent heat being released in the iron snow zone
diffuses to the core mantle boundary. Today's low CMB heat flux
implies that this can be acchieved by a relatively mild temperature
gradient.

\begin{figure*}
\centering
\includegraphics[draft=false,width=0.7\textwidth]{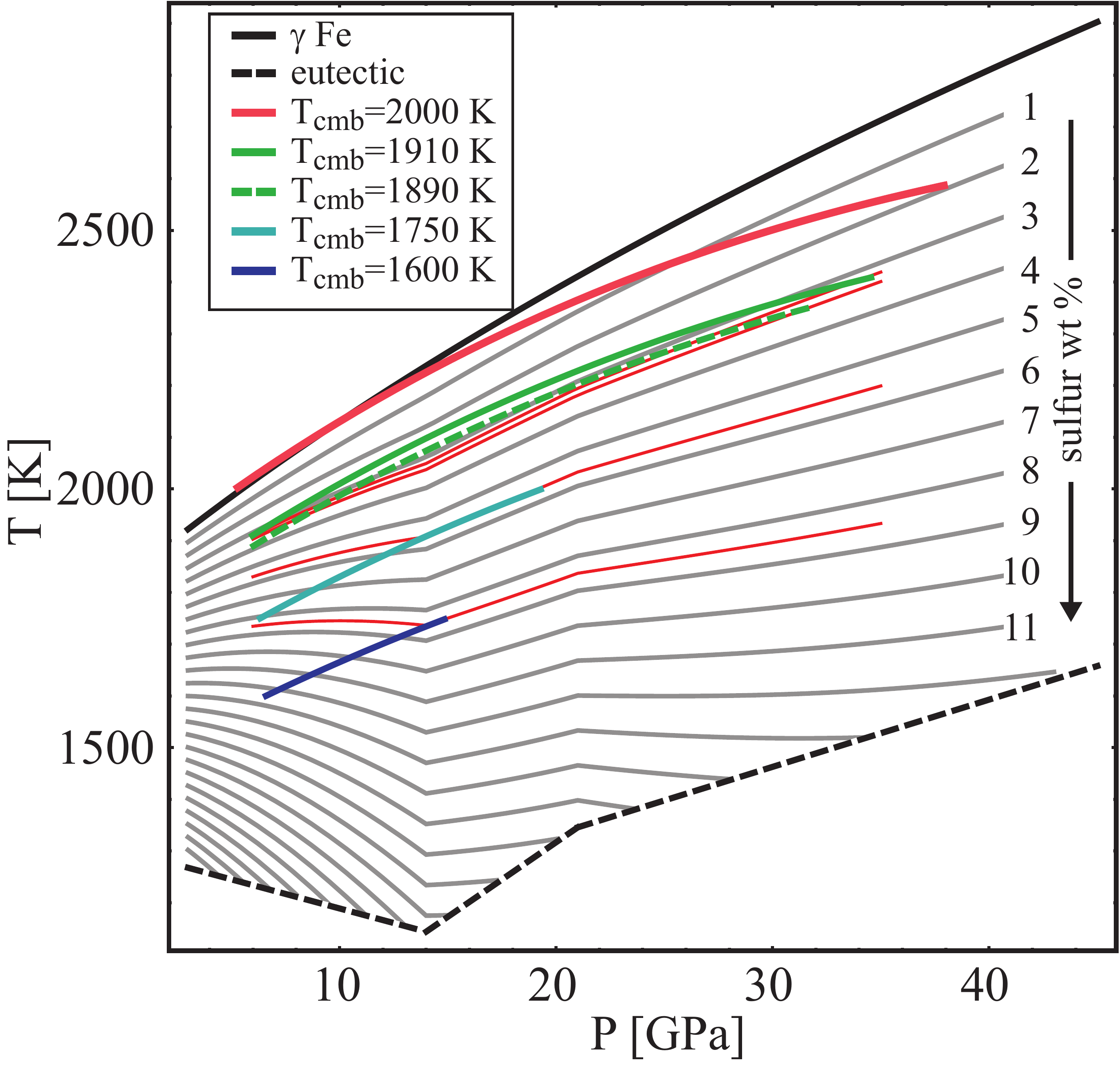}
\caption{Melting curves for different initial sulfur concentrations
and possible Mercury adiabats for different temperatures shown as thick
red, green, turqoise, and blue lines.
Thin red lines from top to bottom show the melting curves for the convecting part of
the core for an initial sulfur concentration of $3\,$wt\% and
a core state described by the solid green, dashed green, gray, and blue
adiabats. The thick solid black line shows the melting curve for pure iron
while the thick dashed black line shows the eutectic temperature.
The figure, provided by Attilio Rivoldini, and has been
adapted from \citet{Rivoldini2011} to include the Mercury core adiabats
calculated in \citet{Rivoldini2013}. A central pressure of $40\,$GPa is
assumed for Mercury but the adiabats are only drawn in the liquid
part of the core.}
\label{Melting}
\end{figure*}

The possible core scenarios are illustrated in
\figref{Melting} with melting curves for different sulfur concentrations and
core adiabats with CMB temperatures in the range between $1600$ and $2000\,$K
suggested by interior \citep{Rivoldini2013} and thermal
evolution models \citep{Grott2011,Michel2013,Tosi2014}.
Data on the melting behaviour of iron-sulfur alloys are
few and the melting curves shown in \figref{Melting} therefore rely on
simple parametrizations \citep{Rivoldini2011}. The adiabats have
been calculated by \citet{Rivoldini2013}.
Mercury's core pressure is only grossly constrained, with CMB pressures
in the range $4-7\,$GPa and central pressures in the range $30-45\,$GPa
\citep{Hauck2013}. We adopt a central pressure
of $40\,$GPa here.
\Figref{Melting} suggests that iron starts to solidify in the
center for an inital sulfur concentrations below about $4\,$wt\%.
Sulfur released from the inner core boundary increases the concentration in
the liquid core over time and thereby slows down the inner core growth
and delays the onset of iron snow.
For an initial sulfur concentration beyond $4\,$wt\% iron
solidification starts with the CMB snow regime.
A convective layer that is enclosed by a solid inner core
and a stably stratified outer iron snow layer seems possible for
sulfur concentrations between about $2.5\,$ and $7\,$wt\%.
For sulfur concentration beyond $7\,$wt\% an inner core
would only grow when the snow zones extends through the whole core
and the snow starts to accumulate in the center.

The adiabats and thin red lines in \figref{Melting} illustrate the
evolution for an initial sulfur concentration of $3\,$wt\%.
For the hot (red) adiabat with $T_{cmb}=2000\,$K neither inner core growth not iron snow
would have started and there would be no dynamo.
When the temperature dops, iron starts to solidify first at the center.
For a CMB temperature of $T_{cmb}=1910\,$K (solid green adiabat),
the inner core has already grown to a radius of about $600\,$km while
the outer snow layer is only about $160\,$km thick.
The sulfur released upon inner core growth has increased the bulk
concentration in the liquide part of the core to $3.4\,$wt\%
(first thin red line from the top).
The decrease in the sulfur abundance due to the remelting of
iron snow has not been taken into account in this model.
When the CMB temperature has dropped to $T_{cmb}=1890\,$K (dashed green adiabat)
the inner core and snow layer have grown by a comparable amount
while the sulfur concentration has increased to $4.4\,$wt\%
(second thin red line from the top).
At $T_{cmb}=1750\,$K (grey) there remains only
a relatively thin convective layer between the inner core boundary at
$r_i=1440\,$km and the lower boundary of the outer snow layer at $r_m=1650\,$km.
For the coldest adiabat shown in \figref{Melting} with
$T_{cmb}=1890\,$K (blue) only the outer $300\,$km of the core remain
liquid but belong to the iron snow zone so that no dynamo seems possible.

Additional sometimes complex scenarios have been discussed in
the context of Ganymede by \citet{Hauck2006} and may also apply at
Mercury since the iron cores of both bodies cover similar pressure ranges.
For example, \figref{Melting} illustrates a kink in the melting curve for
pressures around $21\,$GPa and compositions larger than $5\,$wt\% sulfur.
This could lead to a double snow regime where not only the very outer part
of the core precipitates iron but also an intermediate layer around
$21\,$GPa. This possibility has been explored in a dynamo
model by \citet{Vilim2010} that we will discuss in \secref{Interior}.
Since the kink is not very pronounced, however, such a double snow
dynamo would not be very long lived.

Another interesting scenario unfolds when the light element concentration lies
on the S-rich side of the eutectic. Under these conditions, FeS rather than Fe
would crystalize out when the temperature drops below the FeS melting curve.
Since FeS is lighter then the residuum fluid, the crystals would rise towards
the core-mantle boundary.
However, eutectic or even higher sulfur concentrations cannot represent bulk
conditions since it would be difficult to match Mercury's total mass \citep{Rivoldini2011}.
Inner core growth would increase the sulfur concentration in the remaining
fluid over time but never beyond the eutectic point. This has likely not been reached in Mercury
because the eutectic temperature of $1200-1300\,$K \citep{Rivoldini2011}
is significantly lower than today's CMB temperature suggested by thermal
evolution \citep{Grott2011,Tosi2014} and interior models \citep{Rivoldini2013}.

An alternative explanation for a locally high sulfur concentration
was suggested by the XRS observations. The low Fe but large S abundance in
surface rocks indicates that Mercury's core could have formed at strongly reducing conditions.
This promotes a stronger partitioning of Si into the liquid iron phase
leading to a ternary Fe-Si-S core alloy \citep{Malavergne2010}.
Experiments have shown that Si and S are immiscible for pressures
below $15\,$GPa \citep{Morard2010} which is the pressure range in the
outer part of Mercury's core. However, the immiscibility only happens for
sizable Si and S concentrations. Experiments by \citet{Morard2010}, for example, demonstrate
that at $4\,$GPa and $1900\,$K abundances of $6\,$wt\% S and $6\,$wt\% Si
are required to trigger the immiscibility and lead to the formation of a
sulfur rich phase with a
composition of about $25\,$wt\% S. For FeS crystallization to play a role
at today's CMB temperatures, the sulfur rich phase should lie significantly
to the right of the eutectic where the FeS melting temperature increases
with light element abundance. Thus even higher S and Si contributions
are required but seem once more difficult to reconcile with the planet's
total mass \citep{Rivoldini2013}. Since Si partitions much more easily
into the solid iron phase than sulfur, it's contribution to compositional
convection and the stabilization of the snow zone is significantly
weaker.

Several numerical studies in the context of Earth and Mars
have shown that the CMB heat flux pattern can have a strong effect on the
dynamo mechanism (see e.g.~\citet{Wicht2014} and \citet{Dietrich2013}
for overviews). Like the mean heat flux out of the core, this pattern
is controlled by the lower mantle structure. The Martian dynamo
ceased about $4\,$Gyr ago but has left its trace in form of a strongly
magnetized crust. The fact that the magnetization is much stronger in the
southern than in the northern hemisphere could reflect a special
configuration of the planet's ancient dynamo. Impacts or large degree mantle convection
may have significantly decreased the heat flux through the northern
CMB and therefore weakened dynamo action in this hemisphere
\citep{Stanley2008,Amit2011,Dietrich2013}.
Mercury's magnetic field is distinctively stronger in the northern than in
the southern hemisphere and it seems attractive
to invoke an increased northern CMB heat flux as a possible explanation.

Clues about the possible pattern may once more come
from MESSENGER observations. A combination of gravity and altimeter data allowed to estimate
the crustal thickness in the northern hemisphere. On average, the crust is
about $50\,$km thicker around the equator than around the pole
\citep{Smith2012} which points towards more lava production and thus a hotter
mantle at lower latitudes. This is consistent with the fact that the northern
lowlands are filled by younger flood basalts since melts
more easily penetrate a thinner crust \citep{Denevi2013}.
Missing altimeter data and the degraded precision of gravity measurements
does not allow to constrain the crustal thickness in the southern
hemisphere. The lack of younger flood basalts, however, could
indicate a thicker crust and hotter mantle.
Since a hotter mantle would reduce the CMB heat flux, these ideas
indeed translate
into a pattern with increased flux at higher northern latitudes.
However, Mercury's volcanism ceased more than $3.5\,$Gyr ago and today's
thermal mantle structure may look completely different.
Even simple thermal diffusion should have eroded any asymmetry over
such a long time span.
Thermal evolution simulations show that at least the lower
part of the mantle may still convect today \citep{Smith2012,Tosi2014} which
would change the structure on much shorter time scales.
Since the active shell is so thin, the pattern would be rather small scale
without any distinct north/south asymmetry.

Because of Mercury's 3:2 spin-orbit resonance, the high eccentricity of
the orbit, and the very small obliquity the time averaged insolation
pattern shows strong latitudinal and longitudinal variations.
\citet{Williams2010} calculates that the mean polar temperature can
be $200\,$K lower than the equatorial. Longitudinal variations show
two maxima that are about $100\,$K hotter than the minima at
the equator.
If Mercury's mantle convection has ceased long ago, the respective
pattern may have diffused into the mantle and could determine the CMB
heat flux variation. Higher than average flux at the poles and a somewhat
weaker longitudinal variation would be the consequence.
We discuss the impact of the CMB heat flux pattern on the dynamo
process in \secref{Interior}.

\section{Mercury's external\\ magnetic field}
\label{External}

Planetary magnetospheres are the result of the interaction between the
planetary magnetic field and the impinging solar wind plasma.
Because of Mercury's weak and asymmetric magnetic field and the
position close to the Sun, the Hermean and terrestrial magnetospheres
differer fundamentally.
Mercury experiences the most intense solar wind of all solar-system planets.
Under average conditions, the ratio of the solar wind speed and the Alfv\'en
velocity, called the Alfv\'enic Mach-number, is comparable to the terrestrial
one. With values of $6.6$ for Mercury \citep{winslow_2013} and $8$
for Earth, the solar wind plasma is super-magnetosonic at both planets, i.e.~the medium
propagates faster than magnetic disturbances and a bow shock therefore forms in
front of the magnetosphere.
Because of the weak Hermean magnetic field, the sub-solar point of the bow shock
is located rather close to the planet at an average position of only $1.96$
planetary radii \citep{winslow_2013} compared to $14$ planetary radii for
Earth.

Behind the bow shock, the cold solar wind plasma is heated up
and interacts with the planetary magnetic field, thereby creating the magnetosphere.
To first order, the planetary fieldlines form closed loops within the dayside
magnetosphere and a long tail on the nightside.
The outer boundary of the magnetosphere, the magnetopause, is located
where the pressure of the shocked solar wind and the pressure of the planetary
magnetic field balance. The solar wind ram pressure,
on average $14.3\,$nPa at Mercury \citep{winslow_2013},
is an order of magnitude higher than at Earth while the magnetic field is two
orders of magnitude weaker. Like the bow shock, the magnetopause is therefore
located much closer to the planet at Mercury than at Earth with mean standoff
distances of about $1.45$ \citep{winslow_2013} and $10$ planetary radii,
respectively.
Both the Hermean magnetosphere and magnetosheath, the region between bow shock and
magnetopause, are thus much smaller than the terrestrial equivalents
in relative and absolute terms.

\begin{figure*}
      \centering
      \includegraphics[width=0.6\textwidth]{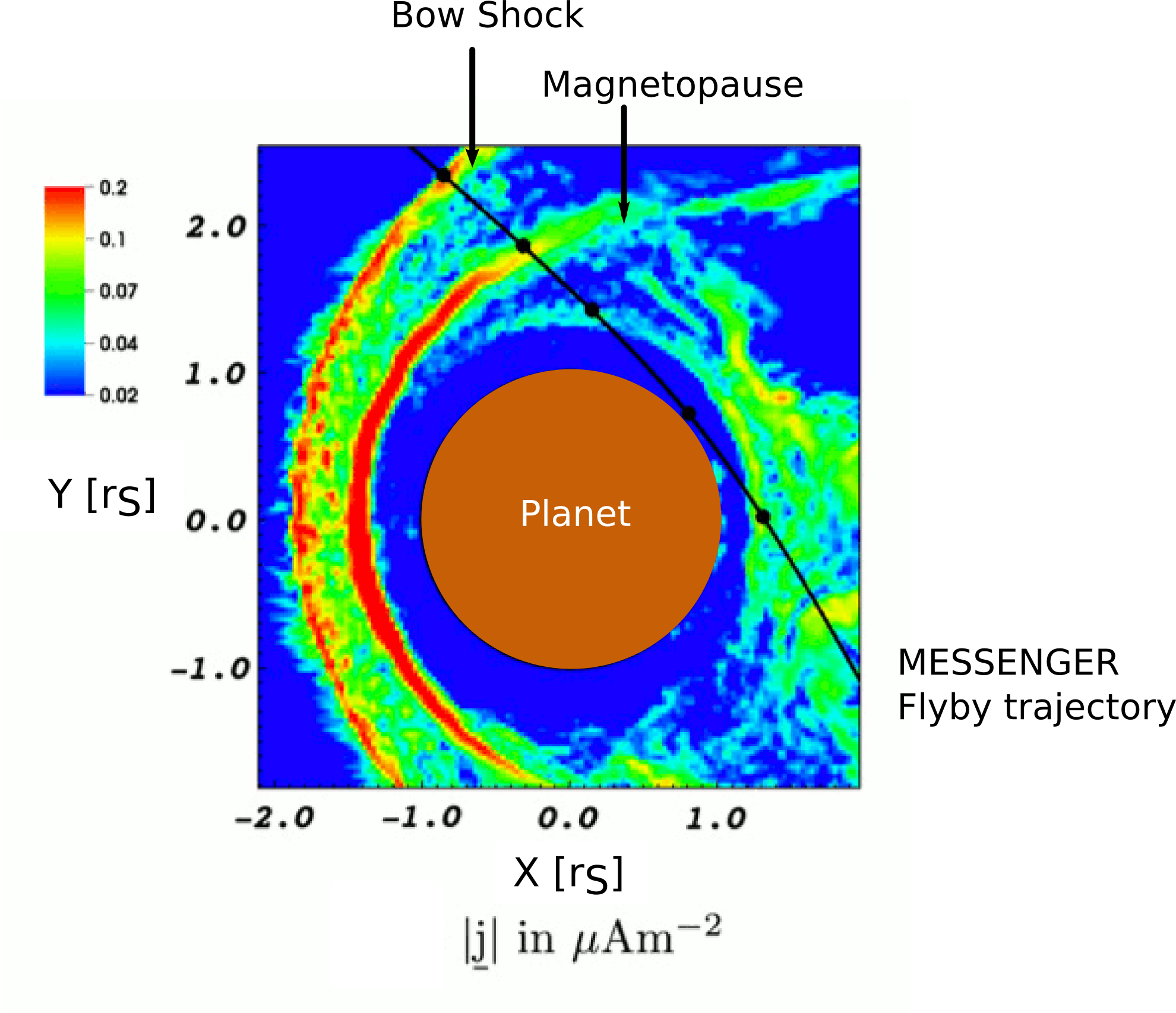}
      \caption{Electrical currents in a numerical simulation of the Hermean magnetosphere.
      The amplitude of the current density $j$ is color-coded.
      An equatorial cross section is shown in an coordinate system where
      $X$ points towards the Sun (negative solar wind direction) and the $Y$-axis
      lies in the Hermean ecliptic. The bow shock standing in front of the planet
      slows down the solar wind.
      The magnetopause is the outer boundary of the magnetosphere.
      The neutral current sheet is located in the nightside of the planet.
      An arc of electrical current visible close to the flyby trajectory (January 14, 2008)
      could be interpreted as a partial ring current.
      This figure is a snapshot from a solar wind hybrid simulation
      and is adapted from \citet{mueller_2012}.
      }
      \label{fig:partial_ring_current_simulation}
\end{figure*}

\begin{figure*}
\centering
\includegraphics[width=0.8\textwidth]{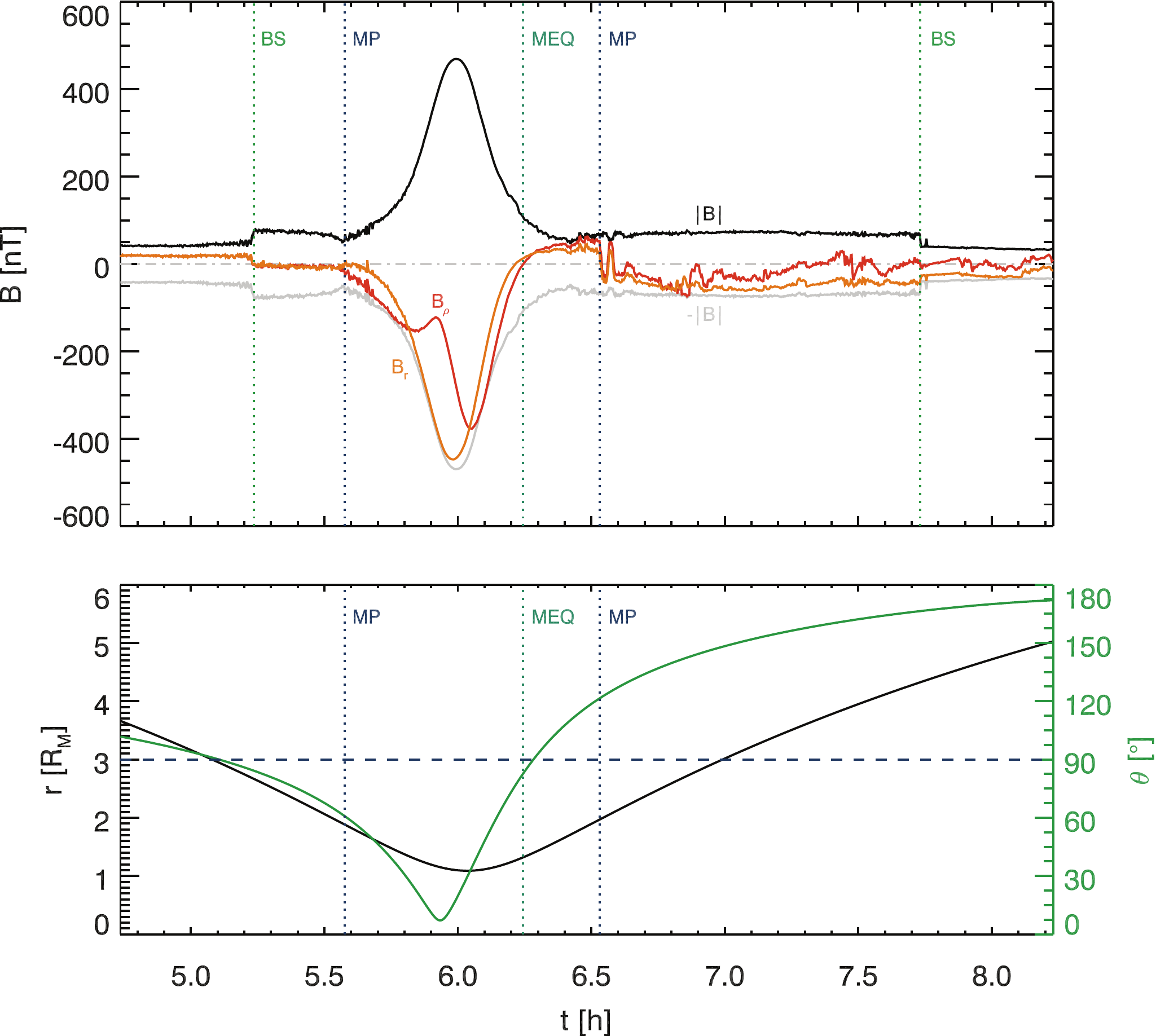}
\caption{Magnetic field data recorded by the MESSENGER magnetometer (10s average)
during orbit $14$ on the DOY $84$ in $2011$. The \textit{upper panel},
shows the time series of the absolute magnetic field $|B|$ (black), the
negative  absolute field (grey), the radial component $B_r$ (green), and
the component $B_\rho$ perpendicular to the rotation axis.
Time is measured in hours since the last apocenter passage.
The plasma boundaries are marked with vertical dashed lines
(BS: bow shock, MP: magnetopause). The location where $B_\rho$ vanishes
defines the magnetic equator (MEQ).
The \textit{lower panel} shows the planetocentric distance $r$ and the co-latitude $\theta$.
The data are taken from the Planetary Data System / Planetary Plasma Interactions Node.}
\label{fig:vis_messenger_orbit}
\end{figure*}

\Figref{fig:partial_ring_current_simulation} shows the current density
in a numerical hybrid simulation that models the solar wind interaction with the
planet \citep{mueller_2012}. The location of the bow shock and the
magnetosphere can be identified via the related current systems.
Along a spacecraft trajectory these boundaries can be
identified by the  related magnetic field changes.
\Figref{fig:vis_messenger_orbit} shows MESSENGER magnetic field measurements
for a relatively quiet orbit (orbit number 14) where both the
bow shock and the magnetopause can be clearly classified on
both sides of the planet.

Another important element of the magnetosphere is the neutral current sheet
which is responsible for the elongated nightside magnetotail and separates the
northern and southern magnetotail lobes. \citet{Johnson2012}
report, that the sheet starts at $1.41\,R_M$, where $R_M$ is the
mean Hermean radius, which is approximately the standoff distance of the
dayside magnetopause. Roughly the same proportion is also found at Earth.

The locations of bow shock, magnetopause and neutral current sheet
is not stationary but vary in time.
The density of the average solar wind decreases with
distance $r_\subtext{S}$ to the Sun like $1 / r_S^2$.
Since Mercury orbits the Sun on a highly elliptical orbit
(ellipticity: $0.21$) the local solar wind pressure varies significantly
on the orbital time-scale of $88\,$days.
The solar wind characteristics also changes constantly on much shorter
time scales because of spatial inhomogeneities due to, for example,
coronal-mass ejections. As a result, the Hermean magnetosphere is very
dynamic. And since the magnetosphere is so small, the
magnetic disturbance also propagate deep into the magnetosphere and
impede the separation of the field into internal and external
contributions \citep{glassmeier_2010}.
Reconnection  processes in the magnetotail are another source
for variations in the Hermean magnetosphere \citep{slavin_2012}.

The Hermean and terrestrial magnetospheres differ in
several additional aspects.
Mercury's surface temperature can reach several hundred Kelvin which
means that the planet's gravitational escape velocity of $4.3$ km/s can
easily be reached thermally.
The thermal escape rate is therefore significant and the remaining atmosphere
too thin to form an ionosphere.
At Earth, the ionosphere hosts substantial current systems that
significantly affect the magnetospheric dynamics, for
example magnetic sub-storms.
Field-aligned currents that close via the ionosphere at Earth must
close within the magnetospheric plasma or the planetary body
at Mercury \citep{janhunen_2004}.

When the planetary magnetic field on the inside of the magnetopause
is nearly antiparallel to the magnetosheath field on the outside,
the respective fieldlines can reconnect.
This typically happens when the interplanetary magnetic field has
a component parallel to the planetary field.
The reconnected fieldlines are advected tail-wards by the solar wind,
which drives a global scale magnetospheric convection loop that ultimately
replenishes the dayside field (Dungey-Cycle).
Due to the small size of the Hermean magnetosphere, the typical timescale of
this plasma circulation is only about $1-2\,$minutes compared
to $1\,$h at Earth \citep{slavin_2012}
which demonstrates that the Hermean magnetosphere can adapt much faster
to changing solar wind conditions.
The rate of reconnection, measured by the relative amplitude of the magnetic
field component perpendicular to the magnetopause, is about $0.15$ at Mercury
and thus $3$ times higher than at Earth \citep{dibraccio_2013}.

\begin{figure*}[t]
\centering
\includegraphics[width=0.6\textwidth]{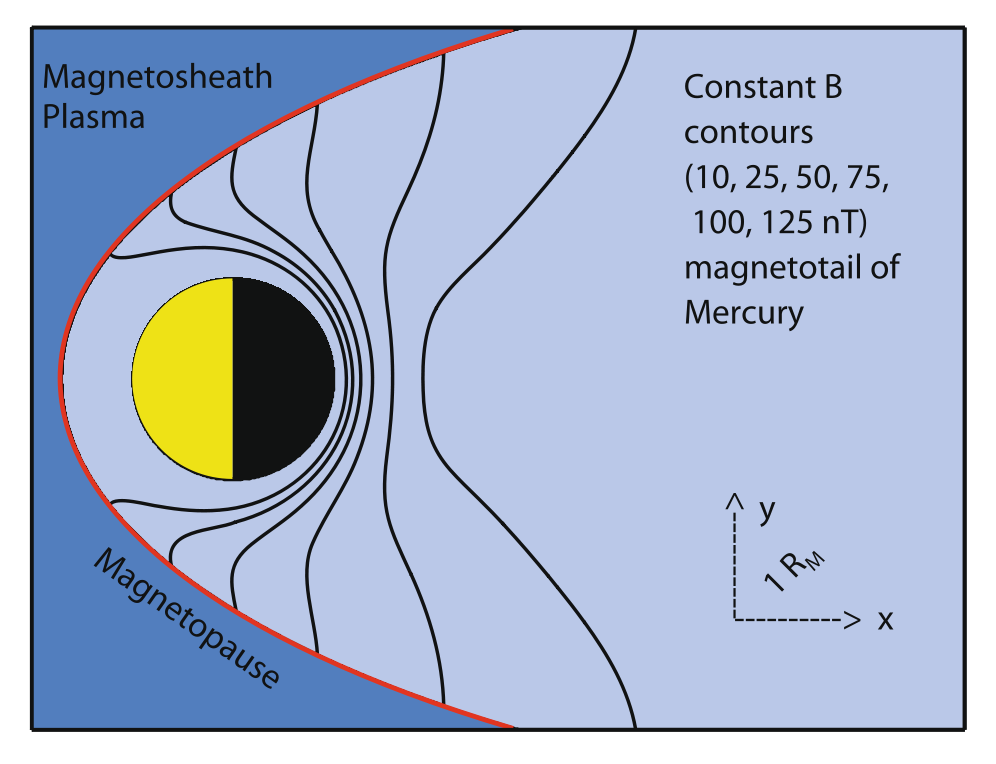}
\caption{Equatorial isocontours of the total magnetic field in a Hermean model magnetosphere.
The magnetopause is shown as a red line and the planet as a sphere.
Figure from \citet{baumjohann_2010}.}
\label{Fig:equatorial_isocontours}
\end{figure*}

Charged particles that are trapped inside the magnetosphere and drift
around the planet in azimuthal direction form a major magnetospheric
current system at Earth, the so-called ring current.
The drift is directed along isocontours of the magnetic field strength.
However, since internal and magnetospheric field can reach comparable values
these contours  close via the magnetopause at Mercury,
as is illustrated in \figref{Fig:equatorial_isocontours}.
At Earth, the planetocentric distance $R_\subtext{rc,E}$ of the ring current
is about four times the terrestrial radius.
When assuming that the position scales linearly with the planetary
dipole moment, the distance can be rescaled to the Hermean situation by
\begin{equation}
      R_\subtext{rc,M} = R_\subtext{rc,E}\,\frac{m_M}{m_E}  \approx 820 \mbox{km} \quad
\end{equation}
where $m_M$ and $m_E$ are the dipole moments of Mercury and Earth, respectively.
The ring current would thus clearly lie below Mercury's surface.
Hybrid simulations by \citet{mueller_2012} indicate that the solar wind protons
entering the magnetosphere can drift roughly half-way around the planet before
being lost to the magnetopause, as is illustrated in
\figref{fig:partial_ring_current_simulation}. This could be interpreted
as a partial ring current. The protons create a
diamagnetic current that locally decreases the magnetic field.

Because MESSENGER delivers only data from one location at a time
inside a very dynamic magnetosphere, it is not only
challenging to separate internal from external field contributions
but also temporal from spatial variations.
Numerical simulations for the solar wind interaction with the planetary magnetic field,
like the hybrid simulation used to investigate the partial ring current
(see \figref{fig:partial_ring_current_simulation}), can improve the
situation by constraining the possible spatial structure for a
given solar wind condition.
However, as these codes are numerically very demanding, it
becomes impractical to perform simulation for all the different
conditions possibly encountered by MESSENGER.
A more practical approach is to use simplified models where a few critical
properties like the shape of the magnetopause and the strength
and shape of the neutral current sheet are described with a few free parameters.
\citet{Johnson2012} demonstrate how the parameters can be fitted
to MESSENGER's accumulated magnetic field data to derive a model for the
time averaged magnetosphere.

\begin{figure*}
      \centering
      \includegraphics[width=0.6\textwidth]{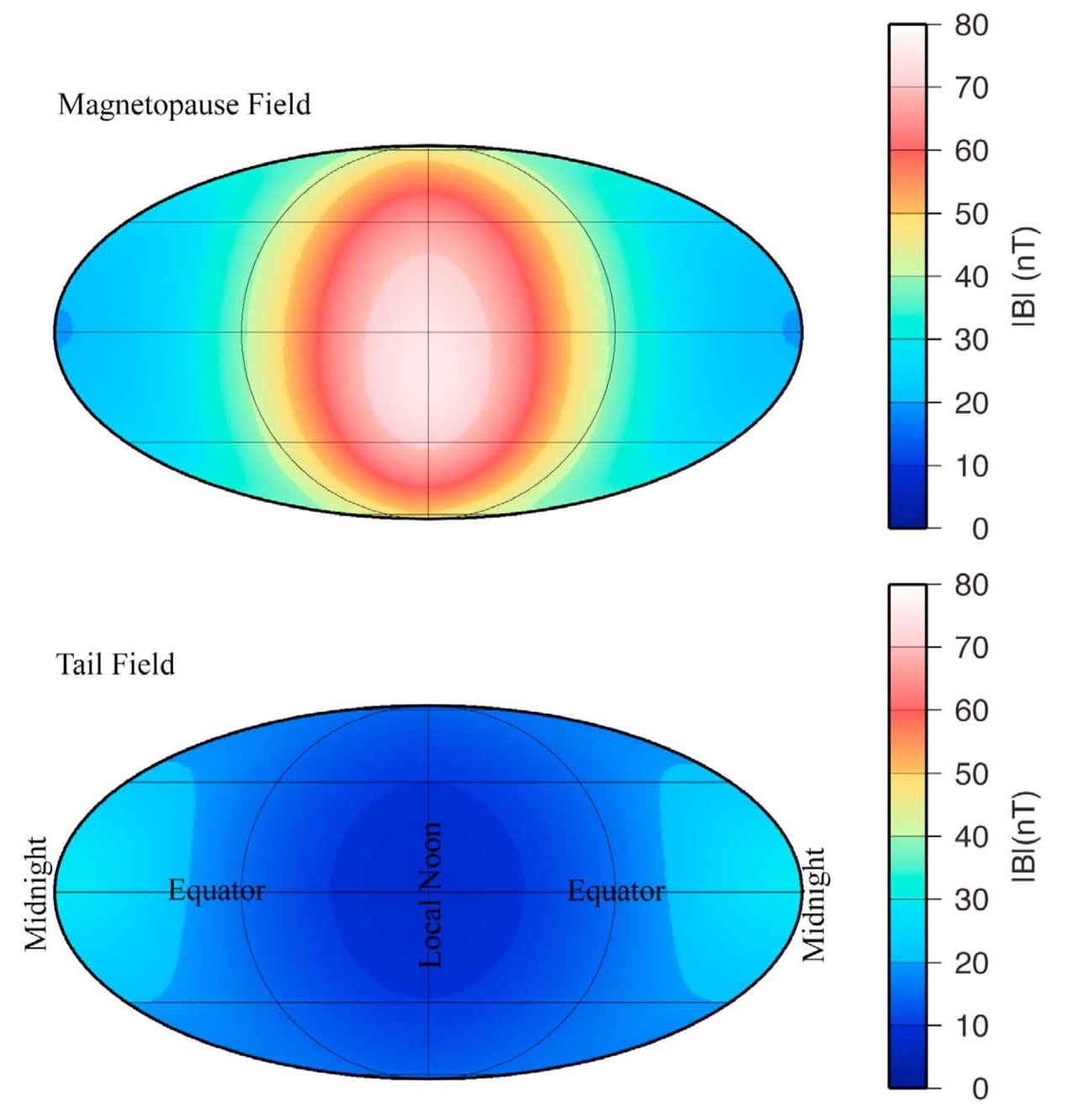}
      \caption{External fields from the paraboloid model based on measurements of
      the MESSENGER mission at the planetary surface. \textit{Top panel:}
      amplitude of the magnetopause field. \textit{Bottom panel:}
      amplitude of the neutral sheet magnetic field.
      Figure from \citet{Johnson2012}.}
      \label{fig:external_field}
\end{figure*}

The offset of Mercury's magnetic field by 20\% of the planetary
radius to the north can cause an equatorial asymmetry of
the planet's exosphere. Ground-based observations of sodium emission lines
suggest that there is more sodium released from the southern
than the northern planetary surface.
\citet{Mangano2013} argue that precipitating solar wind protons are
the main player in the sodium release and more likely reach the
southern surface where the magnetic field is weaker.

The Hermean magnetosphere resembles its terrestrial counterpart in
several aspects but there are also huge differences.
Mercury's magnetosphere is much smaller and significantly more dynamic,
responding much faster to changing solar wind conditions. While the external
field contributions are orders of magnitude smaller than internal
contributions at Earth, they can become comparable at Mercury
(see \figref{fig:external_field}). This lead
\citet{Glassmeier2007} to investigate the long-term
effect of the external field on the internal dynamo process, as we
will further discuss in \secref{Dynamo}.

\section{Mercury's internal\\ magnetic field}
\label{Internal}

The difficulties in separating internal and external field and the
strong covariance of different spherical harmonic contributions caused by the
highly elliptical orbit complicate a classical field modelling with Gaussian
coefficients for Mercury \citep{Anderson2011}. Instead, the MESSENGER
magnetometer team analysed the location of the magnetic equator to indirectly
deduce the internal field. The magnetic equator is the point where the
magnetic field component $B_\rho$ perpendicular to the planetary rotation
axis vanishes. Changing solar wind conditions lead to variations in the
equator location on different time scales from seconds to months but should
average out over time, at least as long as the planetary body itself has no
first order impact on the magnetospheric current system. The mean location of
the magnetic equator is then primarily determined by the internal field.

\citet{Anderson2012} analysed the magnetic equator for 531 descending orbits
with altitudes between $1000$ and $1500\,$km and $120$ ascending orbits with
altitudes between $3500$ and $5000\,$km. They find that the equator crossings
are confined to a relatively thin band offset by about $Z=480\,$km to the
north of the planet's equator. We adopt a planet-centered cylindrical
coordinate system here where $\rho$ and $z$ are the coordinates perpendicular
to and along the rotation axis, respectively, and $\Phi$ is the longitude.
\citet{Anderson2012} minimized the effects of solar wind related magnetic
field variations by considering a mean where each equator location is weighted
with the inverse of the individual standard error $\sigma$. This procedure yields
a mean offset of
$\overline{Z}_{d}=479\,$km with a standard deviation of $\Delta\overline{Z}_d=46\,$km for
the descending orbits. The mean three standard error in determining the
individual equator crossings is $3\overline{\sigma}_d=24\,$km. Because of the
increased solar wind influence and the closer proximity to the magnetosphere,
the magnetic equator is less well defined for the ascending orbits with
$\overline{Z}_{a}=486\,$km, $\Delta\overline{Z}_a=270\,$km, and $
3\overline{\sigma}_a=86\,$km (see table 1 in \citet{Anderson2012}).

\begin{figure*}
\centering
\includegraphics[draft=false,width=0.9\textwidth]{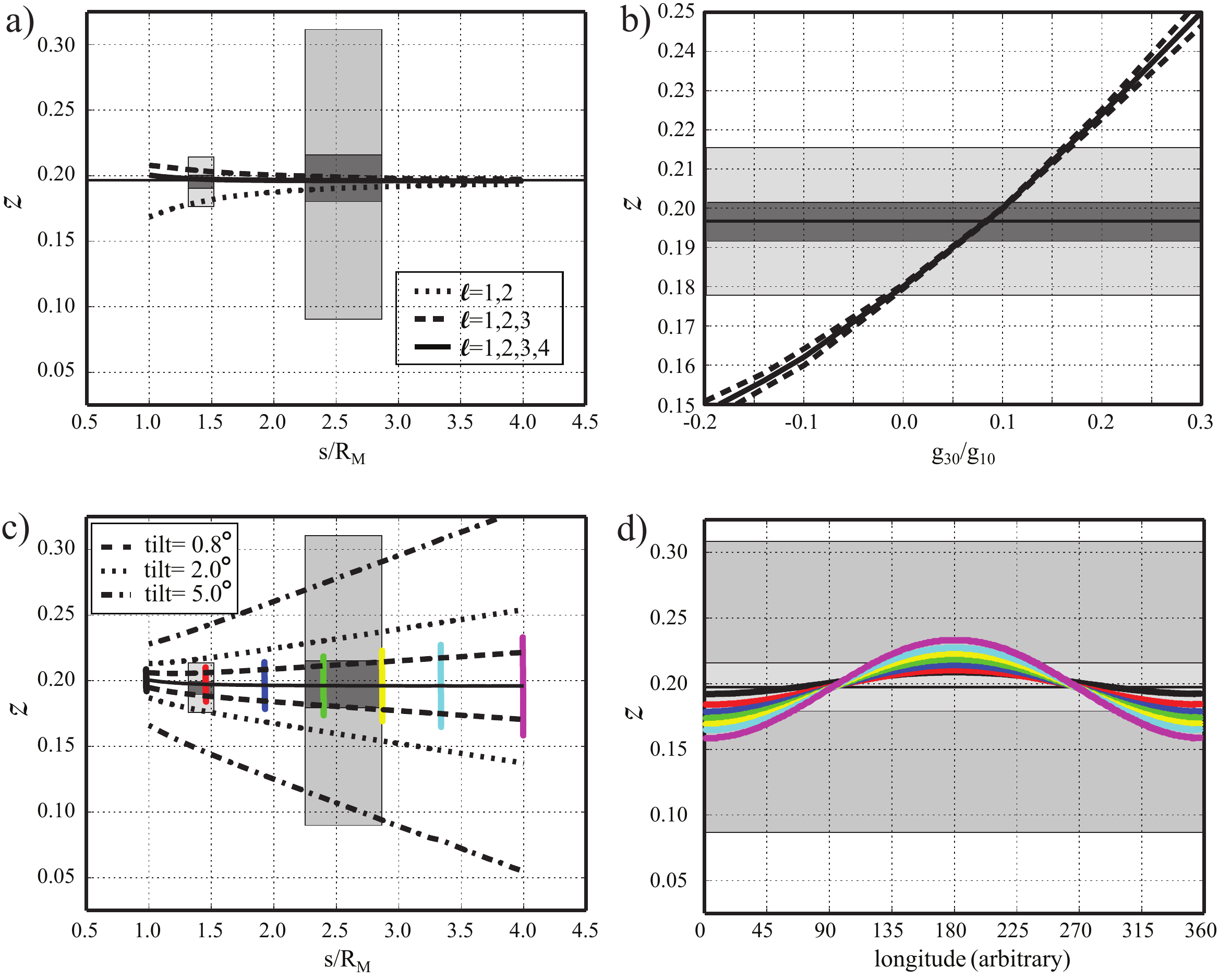}
\caption{Illustration of the offset dipole model by \citet{Anderson2012}.
Panel a) demonstrates how the location of the magnetic equator for
the descending (left box) and ascending (right box) orbits is explained
by combining axial Gauss coefficients up to degree $\ell=4$. Light grey boxes
illustrate the standard deviation, middle grey boxes
the mean three sigma error (see text), and the horizontal black line
corresponds to the mean offset. Panel b) illustrates the impact of
different relative octupole amplitudes $g_{30}/g_{10}$.
Coloured dots in panels c) and d) show the equator locations found on a dense
spherical longitude/latitude grid when an equatorial dipole component $g_{11}$ has been
added that corresponds to a dipole tilt of $0.8^\circ$.
In panel c) the dashed, dotted, and dash-dotted lines show the
mean equator offset for each spherical surface of radius $s/R_M$
plus and minus the standard deviation.}
\label{MEall}
\end{figure*}

These observations suggest that the offset of the magnetic equator has a
constant value of $480\,$km independent of the distance to the planet.
Such a configuration can readily be explained by an internal axial dipole
that is offset by $480\,$km to the North of the equatorial plane.
This translates into an infinite sum of axisymmetric Gaussian
field coefficients $g_\ell$ in the classical planet-centered representation
with
\begin{equation}
\label{OD}
      g_{\ell 0} = \ell\; g_{1 0} \mathcal{Z}^{\ell-1}\;\;,
\end{equation}
where $\mathcal{Z}=Z/R_M$ is the normalized offset and $\ell$ the
spherical harmonic degree \citep{Bartels1936,Alexeev2010}.
Note that all contributions have the same sign.
In the Gaussian representation the planetary surface field is
expanded into spherical surface harmonics $Y_{\ell m}$ of degree $\ell$ and
order $m$ \citep{Olsen2010}. The coefficients $g_{\ell m}$ and $h_{\ell m}$
express the $\cos (m\phi)$ and $\sin (m\phi)$ dependence for
a given degree $\ell$. Only coefficients $g_{\ell 0}$ contribute to
an axisymmetric field.

\citet{Anderson2012} report that coefficients up to $\ell=4$ suffice to
explain the mean magnetic equator locations in the MESSENGER offset
dipole model (MODM). To illustrate the characteristics of MODM,
we experiment with different combinations of the spherical harmonic
contributions and perform a numerical search for the magnetic equator
on a dense longitude/latitude grid for spherical surfaces
with radii up to $4 R_M$. Panel a) in
\figref{MEall} illustrates how the different axisymmetric contributions in
the MODM team up to yield an offset that is nearly independent of the
distance to the planet. A large axial quadrupole contribution which amounts
to nearly $40$\% of the axial dipole guarantees a realistic offset for
$\rho>2 R_M$. Additional higher harmonic contributions are required to
achieve a consistent offset at closer distances. Panel b) in \figref{MEall}
demonstrates that already the relative axial octupole $g_{30}/g_{10}$ is not
particularly well constrained and values between $0.05$ and $0.12$ seem
acceptable. \citet{Anderson2012}, however, suggest a surprisingly tight range of
$0.116\pm 0.009$. Contributions beyond $\ell=3$ can not be particularly
large to retain a nearly constant offset value in the observed range.
Constraining them further, however, would require data closer to the planet
than presently available. The analysis shows that the
mean offset $\mathcal{Z}$ further away from the planet can serve as a proxy
for the ratio of the axial quadrupole to axial dipole contribution while the
dependence of $\mathcal{Z}$ on the distance closer to the planet provides
information on higher order axial contributions.

\citet{Anderson2012} estimate an upper limit for the dipole tilt of
$\Theta=0.8^\circ$. A tilt of the planetary centered dipole causes a
longitudinal variation of the magnetic equator location that increases with
distance to the planet, as is demonstrated in panels b) and c) of
\figref{MEall}. A tilt as large as $2^\circ$ seems still compatible with the
data but the more complex longitudinal dependence of the offset
\citep{Anderson2012} indicates that either higher order harmonics or more
likely the solar wind interaction contributes to the variation around the
mean offset. A tilt below $<0.8^\circ$ is also consistent with a more
complete field analysis by \citet{Johnson2012} that includes a parameterized
magnetospheric model.

\begin{table*}
\centering
\begin{tabular}{cccccc}
 Quantity                & MODM                  & Earth     & Jupiter   & Saturn & Uranus  \\
 \hline
 $g_{10}$ [nT]    &$-190\pm 10$              &$-29\,560$   &$420\,500$&$21\,191$&$11\,855$\\
 tilt [$^\circ$]         & $<0.8$            &$10.2$      &$9.5$     &$<0.06$  &$58.8$\\
 $g_{20}/g_{10}$         &$0.392\pm 0.010$          &$0.079$     &$-0.012$  & $0.075$& $-0.496$\\
 $g_{30}/g_{10}$         &$0.116\pm 0.009$          &$-0.045$    &$-0.004$  & $0.112$& $0.353$\\
 $g_{40}/g_{10}$         &$0.030\pm 0.005$          &$-0.031$    &$-0.040$  & $0.003$& $0.034$\\
 H                       & $0.20$                    & $0.017$    &$0.045$  & $0.050$& $0.251$\\
 \hline
 $\overline{\mathcal{Z}}$           & $2.0\tp{-1}$              &$2.6\tp{-2}$&$3.5\tp{-3}$&$3.8\tp{-2}$ &$5.3\tp{-2}$ \\
  $\overline{\mathcal{Z}}_d$        & $2.0\tp{-1}$              &$3.6\tp{-3}$&$2.8\tp{-2}$&$4.0\tp{-2}$ &$1.6\tp{-1}$ \\
 $\Delta\overline{\mathcal{Z}}$     & $1.7\tp{-2}$$(1.1\tp{-1})$&$2.2\tp{-1}$&$1.6\tp{-1}$&$2.9\tp{-3}$ &$1.0$\\
 $\Delta\overline{\mathcal{Z}}_d$   & $7.5\tp{-3}$$(1.9\tp{-2})$&$1.3\tp{-1}$&$9.3\tp{-2}$&$8.5\tp{-4}$ &$6.6\tp{-1}$\\

\hline
\end{tabular}
\caption{Comparison of Mercury's offset dipole model MODM \citep{Anderson2012}
with magnetic field models for other planets: the Grimm model
\citep{Lesur2012} for Earth, the VIP4 model \citep{Connerney98} for Jupiter,
the model by \citet{Cao2012} for Saturn, and the model by \citet{Holme1996}
up to degree $\ell=4$ for Uranus. The neptunian magnetic field is similar to the
field of Uranus and has therefore not been included. The last four lines list
mean offset values $\overline{\mathcal{Z}}$ for all spherical surfaces up to
$4\,R$ and the mean offset $\overline{\mathcal{Z}}_d$ for the distances between $1.3\,R$ and
$1.5\,R$ covered by MESSENGER's descending orbits. $R$ refers to the planetary
radius ($1\,$bar level for gas planets). $\Delta\overline{\mathcal{Z}}$ and
$\Delta\overline{\mathcal{Z}}_d$ are the related standard deviations. For
Mercury, we list the deviation caused by an $0.8^\circ$ tilt and also the
observed standard deviations in brackets.} \label{TabIF}
\end{table*}

\begin{figure*}
\centering
\includegraphics[draft=false,width=12cm]{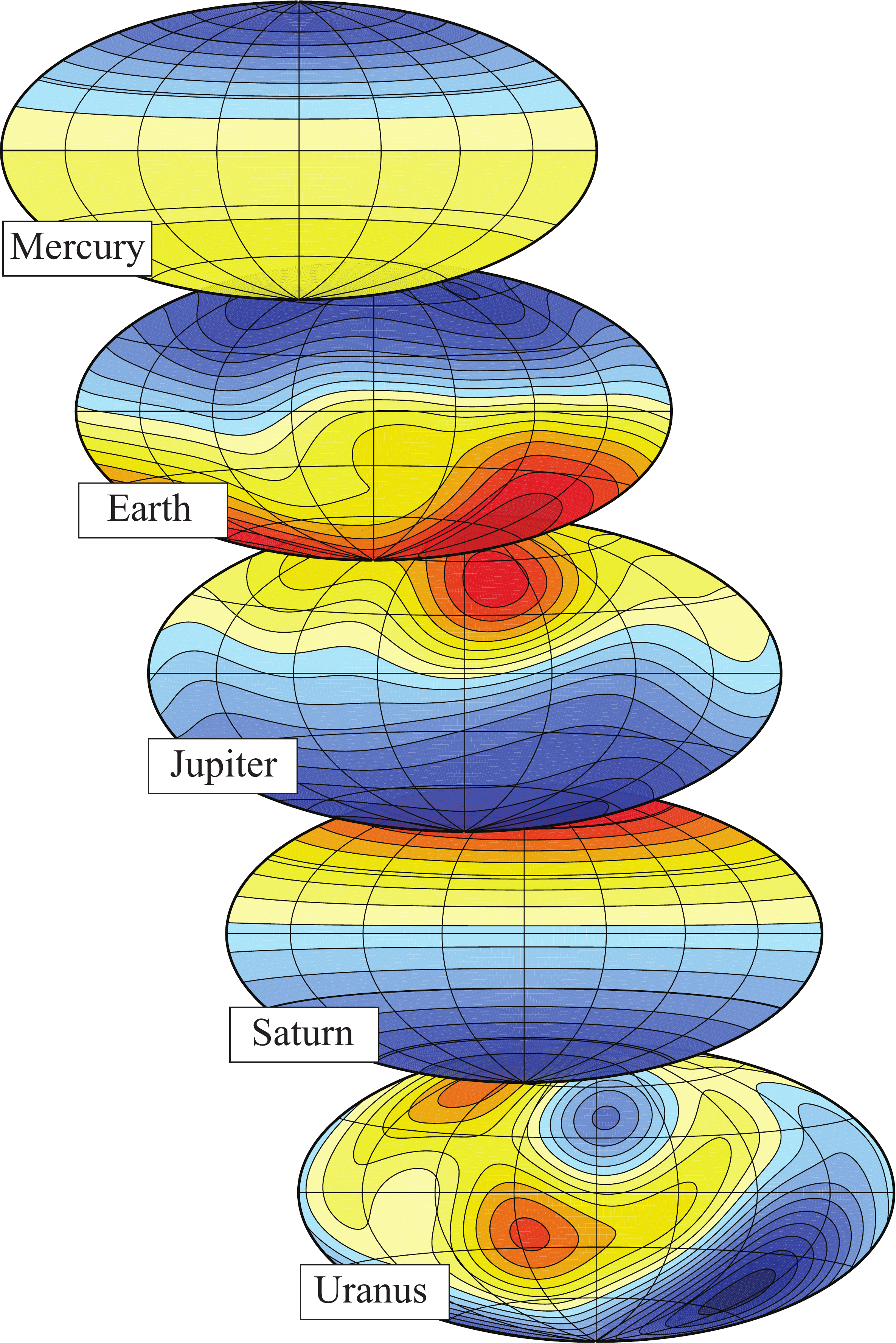}
\caption{Comparison of different radial magnetic fields at
planetary surface. Blue (red and yellow) indicates radially
inward (outward) field. See \tabref{TabIF} for information
on the different field models.}
\label{Comp}
\end{figure*}

\Tabref{TabIF} compares primary magnetic field characteristics of the MODM
with models for other planets and \figref{Comp} shows the respective
radial magnetic surface fields. MODM's large quadrupole contribution is
comparable to that inferred for Uranus or Neptune. Unlike the fields of the ice
giants, however, Mercury's field is also very axisymmetric, a property it
shares with Saturn. The seemingly perfectly axisymmetry of Saturn's field is
also the reason for the small spread $\Delta\overline{\mathcal{Z}}$ of magnetic equator
locations for this planet. Saturn's relative quadrupole contribution, however,
and thus the relative offset is much smaller than at Mercury.

Magnetic harmonics where the sum of degree $\ell$ and order $m$ is odd
(even) represent equatorially anti-symmetric (symmetric) field contributions.
The axial dipole field is thus equatorially anti-symmetric while the
axial quadrupole field is symmetric. Mercury's field has a significant equatorially symmetric
contribution because of the strong axial quadrupole. Another measure
related to the equatorial symmetry breaking is the hemisphericity
\bel{Hemi}
  H= \frac{B_N - B_S}{B_N+B_S}
\ee
where $B_N$ and $B_S$ are the rms surface field amplitudes in the northern
and southern hemispheres, respectively. Due to the offset dipole geometry, the
Hermean magnetic field is significantly stronger in the northern than in the southern
hemisphere so that the hemisphericity reaches a relatively large value of
$0.2$. In conclusion, Mercury's magnetic field is not only very weak but also has a
peculiar geometry unlike any other planet in our solar system that
combines a relatively large axial quadrupole contribution with a very small
dipole tilt.

The time averaged residual field after subtracting the internal and
external field models by \citet{Johnson2012} from the observational
data is surprisingly strong with amplitudes of up to $45\,$nT
at $300\,$km altitude above Mercury's surface \citep{purucker_2012}.
The fact that the residual field is concentrated at high northern latitudes,
is relatively small scale, and correlates
with the boundary of the northern volcanic plains to a fair degree
points towards crustal remanent magnetization, though an internal
field contribution can also not be excluded.
A crustal origin would suggest that Mercury's dynamo is long lived and
probably older than $3.5\,$Gyr. Since the residual field opposes the
current dipole direction, the dynamo must have reversed its polarity
at least once. This would put valuable constraints on thermal
evolution models and dynamo simulations for Mercury.

\section{Modelling Mercury's\\ Internal Dynamo}
\label{Dynamo}

\subsection{Dynamo Theory}

Numerical dynamo simulations solve for convection and magnetic field
generation in a viscous, electrically conducting, and rotating fluid.
Since the solutions are very small disturbances around an adiabatic, well mixed,
non-magnetic, and hydrostatic background state, only first order
terms are taken into account.
For terrestrial planets, the mild density and temperature
variations of the background state are typically neglected in the
so called Boussinesq approximation \citep{Braginsky1995}.
The mathematical formulation of the dynamo problem is then given
by the Navier-Stokes equation
\bea \label{eqn:NS} \E\,\frac{d\U}{d t} &=& -\nabla P - 2\uvz\times\U +
\Ra \frac{r}{r_o} C \uvr  \\&& + \frac{1}{\Pm}(\nabla\times\B)\times\B + \E \nabla^2 \U
\nonumber \;\;, \eea
the induction or dynamo equation
\bel{eqn:Dynamo}
\frac{\partial\B}{\partial t} = (\B\cdot\nabla)\;\U +
\frac{1}{\Pm}\;\nabla^2 \B\;\;, \ee
the codensity evolution equation
\bel{eqn:heat} \frac{d C}{d t} = \frac{1}{\Pra}\,\nabla^2 C + q\;\;,
\ee
the flow continuity equation
\bel{eqn:divU} \nabla\cdot{\U} = 0\;\;,
\ee
and the magnetic continuity equation
\bel{eqn:divB} \nabla\cdot{\B} = 0\;\;. \ee
Here, $d / dt$ stands for the substantial time derivative
$\partial / \partial t + \U\cdot\nabla$, \U\ is the convective flow, \B\ the
magnetic field, $P$ is a modified pressure that also contains centrifugal
effects, and $C$ is the codensity.

The equations are given in a non-dimensional form that uses the
thickness of the fluid shell $d=r_o-r_i$ as a length scale,
the viscous diffusion time $d^2/\nu$ as a time scale, the codensity difference
$\Delta C$ across the shell as the codensity scale, and
$(\rhob\mu\lambda\Omega)^{1/2}$
as the magnetic scale. Here, $r_i$ and $r_o$ are the radii of the
inner and outer boundary, respectively, $\nu$ is the kinematic viscosity,
$\rhob$ the reference state core density, $\mu$ the magnetic permeability,
$\lambda$ the magnetic diffusivity, and $\Omega$ the rotation rate.

The problem is controlled by five dimensionless parameters: the Ekman
number
\bel{eqn:Ek} \E=\frac{\nu}{\Omega d^2}\;\;, \ee
the Rayleigh number
\bel{eqn:Ra} \Ra =
\frac{\bar{g}_o \alpha \Delta c\, d^3}{\kappa\nu} \ee
the Prandtl number
\bel{eqn:Pra}
\Pra=\frac{\nu}{\kappa}\;\;, \ee
the magnetic Prandtl number
\bel{eqn:Pm} \Pm=\frac{\nu}{\lambda}\;\;, \ee
and the aspect ratio
\bel{eqn:ar} a=r_i/r_o\;\;. \ee
These five dimensionless parameters replace the much larger
number of physical properties of which
the thermal and/or compositional diffusivity
$\kappa$, the thermal and/or compositional expansivity $\alpha$,
and the outer boundary reference gravity $\bar{g}_0$ have not been
defined so far.

Convection is driven by density variations due to super-adiabatic temperature
gradients --- only this component contributes to convection --- or due to
deviations from a homogeneous background composition. Possible sources for
thermal convection are secular cooling, latent heat, and radiogenic heating.
Possible sources for compositional convection are the light elements released
from a growing inner core and iron from an iron snow zone.
To simplify computations, both types of density variation are often combined
into one variable called codensity $C$ despite the fact that the molecular
diffusivities of heat and chemical elements differ by orders of magnitude.
The approach is often justified with the argument that the small
scale turbulent mixing, which can not be resolved in the numerical simulation,
should result in larger effective turbulent diffusivities that are of
comparable magnitude \citep{Braginsky1995}. This
has the additional consequence that the `turbulent' Prandtl number and
magnetic Prandtl number would become of order one \citep{Braginsky1995}.
The codensity evolution equation (\ref{eqn:heat}) contains a volumetric
source/sink term $q$ that can serve different purposes depending
on the assumed buoyancy sources. For convection driven by light
elements released from the inner core, $q$ acts as a sink that compensates
the respective source.
When modelling secular cooling, the outer boundary is the sink
and $q$ the balancing volumetric source \citep{Kutzner2000}.
For iron snow that remelts at depth
$q$ should be positive in the snow zone but negative in the convective
zone underneath.

Typically, no-slip boundary conditions are assumed for the flow.
For the condensity, either fixed codensity or fixed
flux boundary conditions are used. The latter translates
to a fixed radial gradient and requires a modification of the
Rayleigh number \refp{eqn:Ra} where $\Delta C$ then stands for the imposed
gradient times the length scale $d$. For terrestrial planets, the much slower
evolving mantle controls how much heat is allowed to leave the core, so that
a heat flux condition is more appropriate. Lateral variations on the thermal
lower mantle structure translate into an inhomogeneous
core-mantle boundary heat flux \citep{Aubert2008}.
Since the electrical conductivity of the rocky mantle in terrestrial planets is orders
of magnitudes lower than that of the core, the magnetic field can be assumed
to match a potential field at the interface $r=r_o$. This matching condition
can be formulated as a magnetic boundary condition for the individual
spherical harmonic field contributions \citep{Christensen2007}.
A simplified induction equation \refp{eqn:Dynamo} must be solved for the
magnetic field in a conducting inner core which has to match the outer
core field at $r_i$. We refer to \citet{Christensen2007} for a more detailed
discussion of dynamo theory and the numerical methods employed to
solve the system of equations.

Explaining the weakness of Mercury's magnetic field proved a challenge
for classical dynamo theory. In convectively driven core dynamos, the Lorentz force
and thus the magnetic field needs to be sufficiently strong to influence the
flow and thereby saturate magnetic field growth.
The impact of the Lorentz force is often expressed via the
Elsasser number
\bel{Els}
\Lambda=B^2 /\rho\mu\lambda\Omega
\ee
where
$B$ is the typical magnetic field strength.
The Elsasser number estimates the ratio
of the Lorentz to the Coriolis force which is known to enter the
leading order convective force balance. For Earth, $\Lambda$ is
of order one which suggests that the Lorentz force is indeed significant.
For Mercury, however, extrapolating the measured surface field strength
to the planet's core mantle boundary yields $\Lambda_{cmb}\approx 10^{-5}$,
a value much too low to be compatible with an Earth-like convectively
driven core dynamo \citep{Wicht2007}.
Several authors therefore pursued alternative theories like crustal
magnetization or a thermo-electric dynamo
(see \citet{Wicht2007} for an overview).

However, convectively driven core dynamos remain the preferred explanation
since different modifications of the numerical models originally developed to
explain the geodynamo successfully reduced the surface field
strength towards more Mercury like values
(for recent overviews see \citet{Wicht2007,Stanley2010,Schubert2011}).
We revisit several of these models in the following and test whether
they are consistent with MESSENGER magnetic field data.

\begin{landscape}
\begin{table}
{\scriptsize
\begin{tabular}{c|c|ccc|ccc|ccc}
\hline
Model          & MODM   & E5R6 & E5R36 & E5R45 & CW2 & CW3 & CW4 & $Y_{10}$ BD & $Y_{10}$ ID & $Y_{20}$ \\
\hline
\Ra\           & --- &$2\tp{7}$&$1.2\tp{8}$&$1.5\tp{8}$&$2\tp{8}$&$4\tp{8}$&$6\tp{8}$&$4\tp{7}$&$4\tp{7}$&$4\tp{7}$\\
\E\            &$10^{-13}$ &$3\tp{-5}$&$3\tp{-5}$&$3\tp{-5}$&$10^{-4}$&$10^{-4}$&$10^{-4}$&$10^{-4}$&$10^{-4}$&$10^{-4}$\\
\Pm\           &$10^{-6}$  &  $1$ &$1$ &$1$ &$3$&$3$&$3$&$2$&$2$&$2$\\
\Pra\          &$0.1$      &$1$&$1$&$1$&$1$&$1$&$1$&$1$&$1$&$1$\\
$a$            & ---       &$0.35$&$0.35$&$0.35$&$0.50$&$0.50$&$0.50$&$0.35$&$0.35$&$0.20$\\
\hline
\Rol\          &  8        &$0.02$&$0.10$&$0.18$&$0.42$&$2.7$&$3.6$&$0.11$&$0.06$&$0.05$\\
$\Lambda_{cmb}$&$10^{-5}$  &$2.2\tp{-2}$&$1.8\tp{-1}$&$1.6\tp{-2}$&$4.9\tp{-4}$&$1.9\tp{-4}$&$4.7\tp{-5}$&$1.8\tp{-1}$&
$1.9\tp{-1}$&$4.0\tp{-2}$\\
$|g_{10}|$[nT] &$190$&$8.8\tp{3}$&$2.6\tp{4}$&$1.4\tp{3}$&$1.4\tp{3}$&$924$&$432$&$2.1\tp{4}$&
$1.7\tp{4}$&$8.3\tp{3}$\\
tilt [$^\circ$] &$<0.8$& 0 &$2.5$&$38.1$&$3.5$&$4.2$&$8.6$&$3.4$&$10.7$&$8.2$\\
$|g_{20}/g_{10}|$&$0.39$     & 0 &$0.34$&$3.8$&$0.08$&$0.16$&$0.31$&$0.06$&$0.25$&$0.52$\\
$H$            &$0.20$     & 0 &$0.02$&$0.11$&$0.05$&$0.09$&$0.09$&$0.04$&$0.20$&$0.23$\\
$|\overline{\mathcal{Z}}|$&$2.0\tp{-1}$&0&$2.2\tp{-2}$&$2.1\tp{-1}$&$4.0\tp{-2}$&$8.1\tp{-2}$&$8.4\tp{-2}$&$3.8\tp{-2}$&
$1.5\tp{-1}$&$2.0\tp{-1}$\\
$\Delta\overline{\mathcal{Z}}$&$1.7\tp{-2}\;(1.1\tp{-1})$&0&$6.9\tp{-2}$&$8.6\tp{-1}$&$8.3\tp{-2}$&$2.4\tp{-1}$&$1.1\tp{-1}$&
$2.2\tp{-1}$&$2.3\tp{-1}$\\
\hline
\end{tabular}
}
\caption{Comparison of Mercury's dynamo parameter and properties with
nine dynamo models that have been (re)analysed here. For Mercury we
list the MODM while time averaged
local Rossby numbers $\Rol$ and magnetic field properties are listed for
the numerical simulations. Mercury's core Rayleigh number and aspect ratio
are basically unconstrained but and we have assumed an Earth-like value
of $a=0.35$ to calculate the Ekman number.
Other Mercury parameters follow \citet{Schubert2011} and
\citet{Olson2006b}. Two models with a $Y_{10}$ CMB heat flux patter are listed, one is
bottom driven (BD) while the other is internally driven (ID). The model
with a $Y_{20}$ pattern is internally driven.}
\label{TabPar}
\end{table}
\end{landscape}

All dynamo simulations have the problem that numerical
limitations prevent the use realistic diffusivities. For example,
the viscous diffusivity is many orders of magnitude too large to
damp the very small scale convection motions that cannot be resolved
with the available computer power. Dynamo modelers typically
fix the Ekman number $\E$, the ratio of viscous to Coriolis forces, to
the smallest value accessible with the numerical resources.
The most advanced computer simulations reach down to $E=10^{-7}$ which
is still many orders of magnitude larger than the planetary value of
$\E\approx 10^{-12}$ (see \tabref{TabPar}). The Prandtl number \Pra\ can
assume realistic values but the magnetic Prandtl number $\Pm$ has to
be set to a value that guarantees dynamo action. Because of the increased
viscous diffusivity, $\Pm$ is also orders of magnitudes too large.
The Rayleigh number is then adjusted to a value that yields the desired
dynamics. The fact that numerical Dynamo simulations are very successful
in reproducing many aspects of planetary dynamos suggest that at least
the large scale dynamics responsible for producing the observable magnetic
field is captured correctly.

The simulation results must be rescaled to the planetary situation.
For simplicity, we will rescale
the magnetic field strength by assuming that the Elsasser number
would not change when pushing the parameter towards realistic values.
Assuming Mercury's rotation rate, mean core density, magnetic permeability
and magnetic diffusivity then allows the deduce the dimensional
magnetic field strength via \eqnref{Els}. Note, however, that
other scalings have been proposed \citep{Christensen2010} and may
lead to somewhat different answers.

\subsection{Standard Earth-like Dynamo Models}

To highlight the difficulties of classical dynamo simulations to reproduce
the Hermean magnetic field, we start with analysing three models that
have been explored in the geomagnetic context by \citet{Wicht2010b}.
All have the same Ekman ($\E=3\tp{-5}$), Prandtl number ($\Pra=1$),
magnetic Prandtl number ($\Pm=1$), aspect ratio ($a=0.35$), use rigid and
fixed codensity boundary conditions and are driven by a growing inner core.
They differ only in the Rayleigh number: Model E5R6 has the
lowest Rayleigh number of $\Ra=2\tp{7}$, six times the critical
value for onset of convection. Model E5R36 has an intermediate
Rayleigh number of $\Ra=1.2\tp{8}$ while model E5R45 has the
largest Rayleigh number at $\Ra=1.5\tp{8}$.
All model parameters are listed in \tabref{TabPar}.

\begin{figure*}
\centering
\includegraphics[draft=false,width=0.9\textwidth]{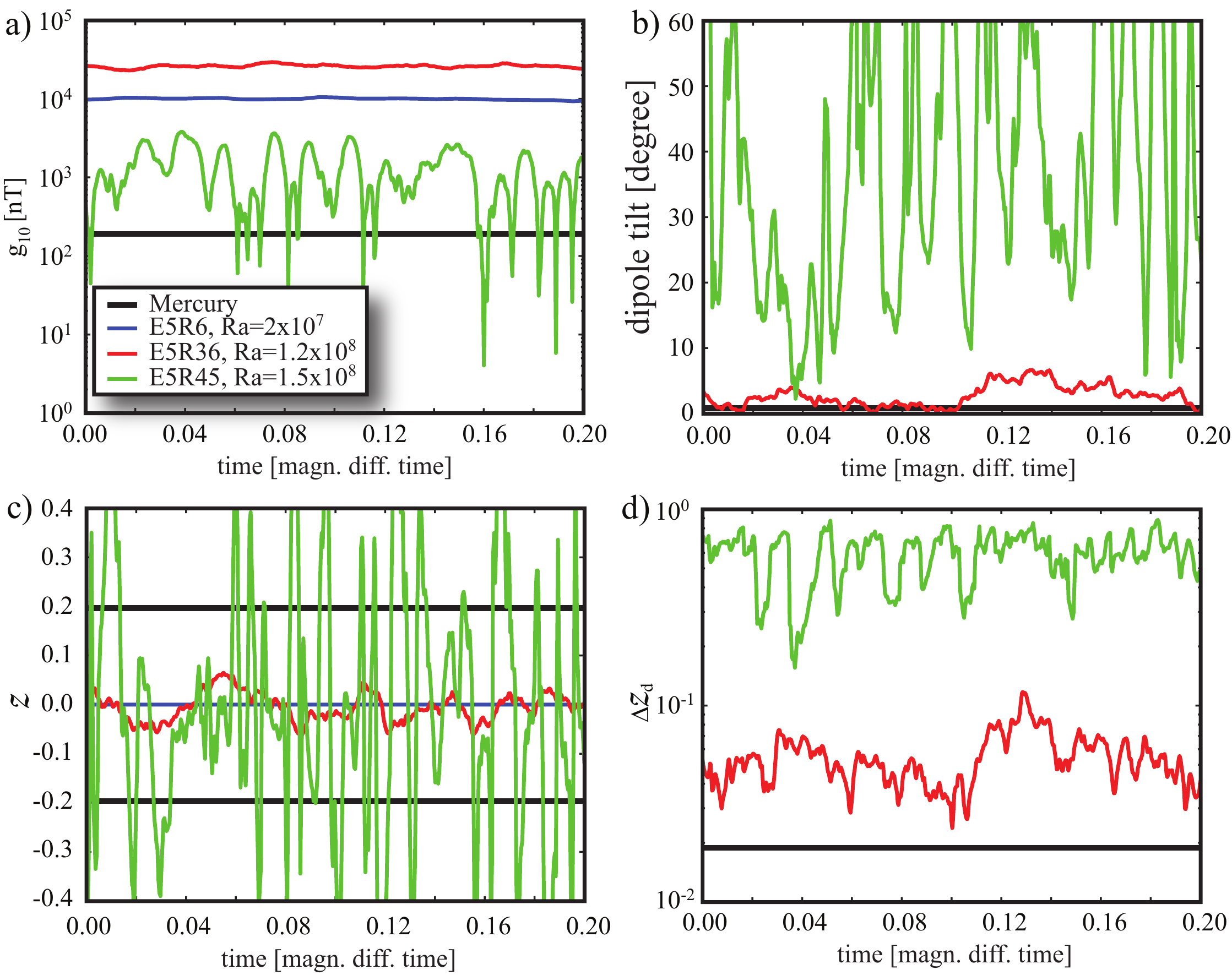}
\caption{Time evolution of three standard dynamo models with
different Rayleigh numbers. The thick black horizontal lines
indicate the MESSENGER offset dipole model.
Panel a) shows the axial dipole coefficient, panel b) the dipole tilt,
panel c) the mean offset $\mathcal{Z}$ averaged over all radii
up to $4 R_M$, and panel d) shows the standard deviation
for the offset in the distance range of the descending
orbits $\Delta\mathcal{Z}_d$. For the numerical simulations
time is given in units of the magnetic diffusion time
$\tau_\lambda=d^2/\lambda$. When assuming an Earth-like
aspect ratio of $0.35$ and a magnetic diffusivity of
$\lambda=1$ the Hermean magnetic diffusion
time amounts to $\tau_\lambda\approx 54\,$kyr.}
\label{MerT_BH}
\end{figure*}

\Figref{MerT_BH} shows the time evolution of the axial dipole strength,
the dipole tilt, the mean magnetic equator offset $\overline{\mathcal{Z}}$
for up to four planetary radii (assuming Mercury's thin crust),
and the related standard deviation $\Delta\overline{\mathcal{Z}}$.
At the lowest Rayleigh number, convective driving is too small
to break the equatorial symmetry. The magnetic field is therefore
perfectly equatorially anti-symmetric and very much dominated by the axial
dipole contribution. Dipole tilt, offset and standard deviation therefore vanish.
At the intermediate Rayleigh number the solution is sufficiently dynamic and
asymmetric to be considered very Earth-like
\citep{Wicht2010b,Christensen2010a}. While the axial quadrupole and other
equatorially symmetric field contributions have grown, the strong axial dipole
still clearly dominates. The mean offset $\overline{\mathcal{Z}}$ therefore remains small
but oscillates around zero since neither the northern nor the
southern hemisphere are preferred in the dynamo setup.
The spread $\Delta\overline{\mathcal{Z}}$ is of the same
order as the offset itself mainly because of the Earth-like dipole tilt.
The inertial contributions in the flow force balance have increased to a point
where magnetic field reversals can be expected \citep{Christensen2006,Wicht2010b}.

\citet{Christensen2006} introduced the local Rossby number
\bel{Rol}
\Rol=\frac{\mbox{U}}{L \Omega}
\ee
to quantify the ratio of inertial to Coriolis forces. Here, U is the rms
flow amplitude and $L$ is a typical flow
length scale defined by
\bel{ell} L = d\,\pi\,\frac{\sum \mbox{U}_\ell}{\sum \ell \mbox{U}_\ell}\;\;.
\ee
U$_\ell$ is the rms flow amplitude of
spherical harmonic contributions with degree $\ell$.
The Coriolis force is responsible for organizing the flow into
quasi two-dimesional convective columns which tend to produce
the larger scale dipole dominated magnetic field. Inertia and in
particular the non-linear advective term, on the other hand, is
responsible for the mixing of different scales and therefore the braking of
flow symmetries.
At $\Rol=0.10$ inertia is likely large enough in model E5R36
to trigger reversals
though no such event has been observed in the relatively short period
we could afford to simulate.

 \citet{Christensen2006} report
that this typically happens for $\Rol\approx 0.1$, a limit clearly exceeded
at $\Rol=0.18$ in model $E5R45$.
Smaller scale contributions dominate the now multipolar
magnetic field which also becomes very variable in time
and constantly changes its polarity \citep{Wicht2010b}. Consequently, the offset
also varies rapidly and may even exceed Mercury's offset value at times where
the axial dipole is particularly low (see \figref{MerT_BH}b).
While axial dipole and offset can assume Mercury-like values during brief
periods in time, this is not true for the dipole tilt and
$\Delta\overline{\mathcal{Z}}$.
The larger Rayleigh number promotes not only the axial quadrupole
but higher harmonics and non-axial field contribution in general.
The tilt is therefore typically rather large and the magnetic equator
covers a wide latitude range. Closer to the planet,
even two or more closed lines with $B_\rho=0$ can be found at a given radius.

\citet{Olson2006b} estimate a large local Rossby number of
$\Rol\approx8$ for Mercury, mainly because of the planet's
slow rotation rate. This suggests that the dynamo produces
a multipolar field at least as complex as in the large Rayleigh number
model E5R45. This is at odds with the observations unless
we could add a physical mechanism to the model that would
filter out smaller scale field contributions while retaining
the strong axial quadrupole. As we discuss in the following,
the stably stratified layer underneath Mercury's core-mantle
boundary \secrefp[See]{Interior} may meet these requirements.

\subsection{Dynamos with a stably stratified outer layer}

The idea of a stably stratified layer in the outer part of a dynamo
region was first proposed by \citet{Stevenson1980} to explain Saturn's
very axisymmetric magnetic field.
The immiscibility of Helium and Hydrogen in Saturn's metallic envelope
\citep{Lorenzen2009} may cause Helium to precipitate into the
deeper interior. Similar to the iron snow scenario discussed in
\secref{Interior} this process may establishes a stabilizing Helium
gradient in the rain zone.

\citet{Christensen2006b} and \citet{Christensen2008} adopt this idea for
Mercury. They propose that the subadiabatic heat flux through the CMB
leads to the stable stratification but since they use a condensity
approach the model is not able to distinguish between thermal
and compositional effects.
The magnetic field that is produced in the convecting deeper core region has
to diffuse through the largely stable outer layer so that the
magnetic skin effect applies here.
The time variability of the magnetic field increases with spatial
complexity \citep{Christensen2004,Lhuillier2011}. The higher harmonic field contributions
are therefore more significantly damped by the skin effect than for example
dipole or quadrupole. Zonal motions that may still penetrate the
stable layer cannot lead to significant dynamo action but further
increase the skin effect for non-axisymmetric field contributions.
Thanks to this filtering effect, the multipolar field of
a high $\Rol$ dynamo should look more Mercury-like when reaching
the planetary surface.

Testing different dynamo setups, \citet{Christensen2006b} and \citet{Christensen2008}
demonstrate that the surface field is indeed weaker and less complex when
a sizable stable layer is included.
We reanalyse the models 2,3, and 4 published in \citet{Christensen2008}
to test whether they are consistent with the new MESSENGER data.
All three models, that we will refer to as CW2, CW3, and CW4 in the
following, have a solid inner core that occupies the inner 50\% in
radius and a stable region that occupies the outer 28\%.
Thus only a relatively thin region is left to host the active dynamo.
Like for the standard models explore above,
all three cases have the same Ekman number $(\E=10^{-4})$,
Prandtl number $(Pr=1)$, and magnetic Prandtl number $(Pm=3)$ but differ
in Rayleigh number. Once more, they use rigid flow boundary conditions
and are driven by a growing inner core.
The model parameters are listed in \tabref{TabPar}.

\begin{figure*}
\centering
\includegraphics[draft=false,width=0.9\textwidth]{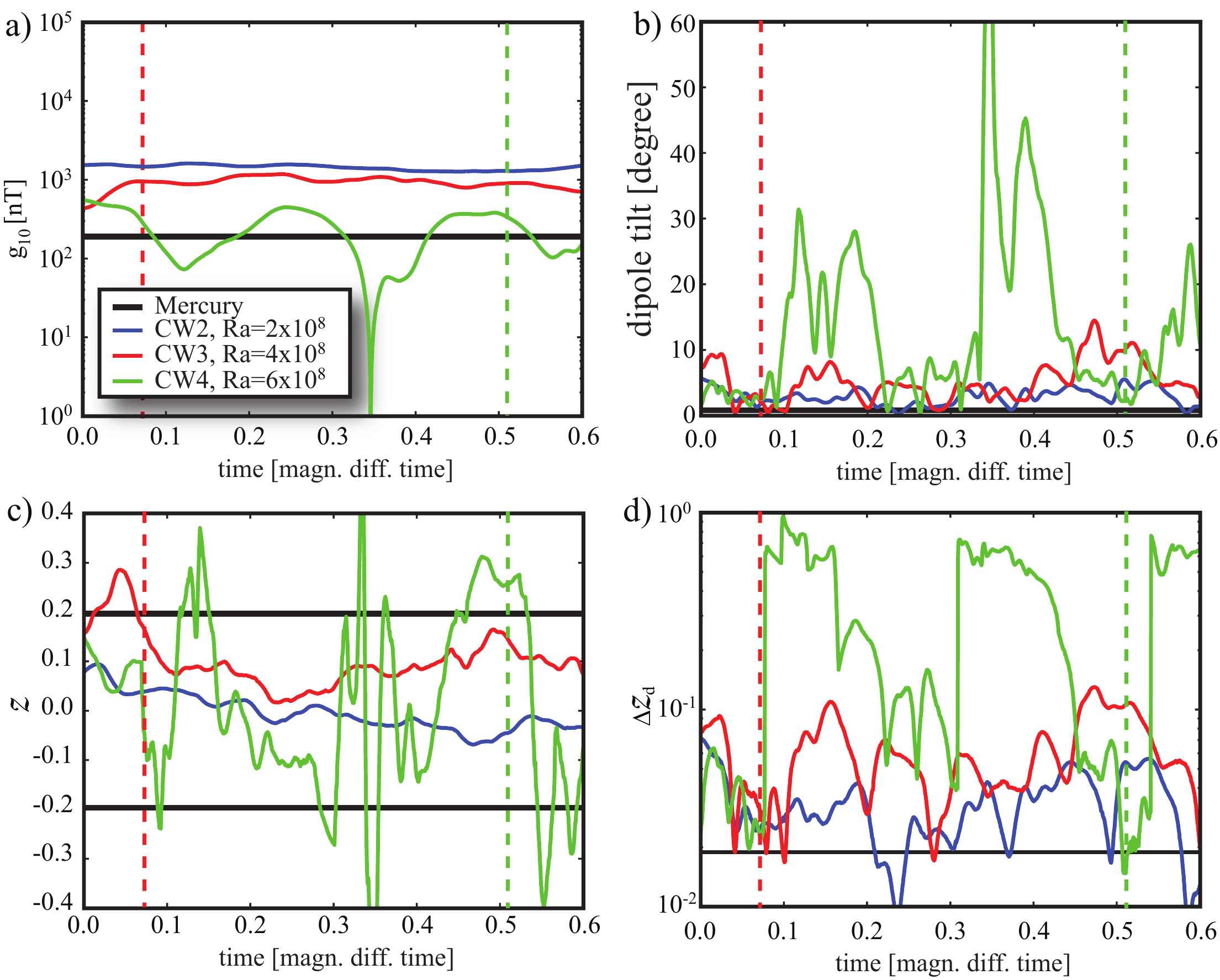}
\caption{Time evolution of three dynamo models with
a stably stratified layer. See \figref{MerT_BH} for more explanation.
The dashed vertical red and green lines marke the times for the
snapshots illustrated in \figref{MEall_Uli} and \figref{Br_Sur_CW}.}
\label{MerT_Uli}
\end{figure*}

\Figref{MerT_Uli} shows the time evolution of the axial dipole contribution,
the dipole tilt, the mean offset $\mathcal{\overline{Z}}$, and of its standard
deviation
$\Delta\mathcal{\overline{Z}}$. At the lowest Rayleigh number of $\Ra=2\tp{8}$ in CW2, the
magnetic field strength is already significantly weaker and the dipole
tilt and offset standard deviation can actually reach Mercury-like small values.
The axial dipole component, however, is still somewhat strong and dominant
and the offset value therefore too small.
Increasing the Rayleigh number to $\Ra=4\tp{8}$ in model CW3 decreases the axial dipole
in absolute and relative terms. The mean tilt, offset, and spread increase,
but there are times when Mercury-like field geometries are approached.
The axial dipole is still by a factor four too strong.
At $\Ra=4\tp{8}$ in model CW4, however, the axial dipole can even become
smaller than at Mercury. The field is very time dependent during
these episodes and is characterized by large dipole tilts and
$\Delta\mathcal{\overline{Z}}$ values since higher harmonic and non-axisymmetric
field contributions dominate.
Very Mercury-like fields, that combine small dipole tilts with larger offset values
but small offset standard deviations, can be found during brief periods
when the axial dipole is somewhat stronger than the Mercury value.

\begin{figure*}
\centering
\includegraphics[draft=false,width=0.9\textwidth]{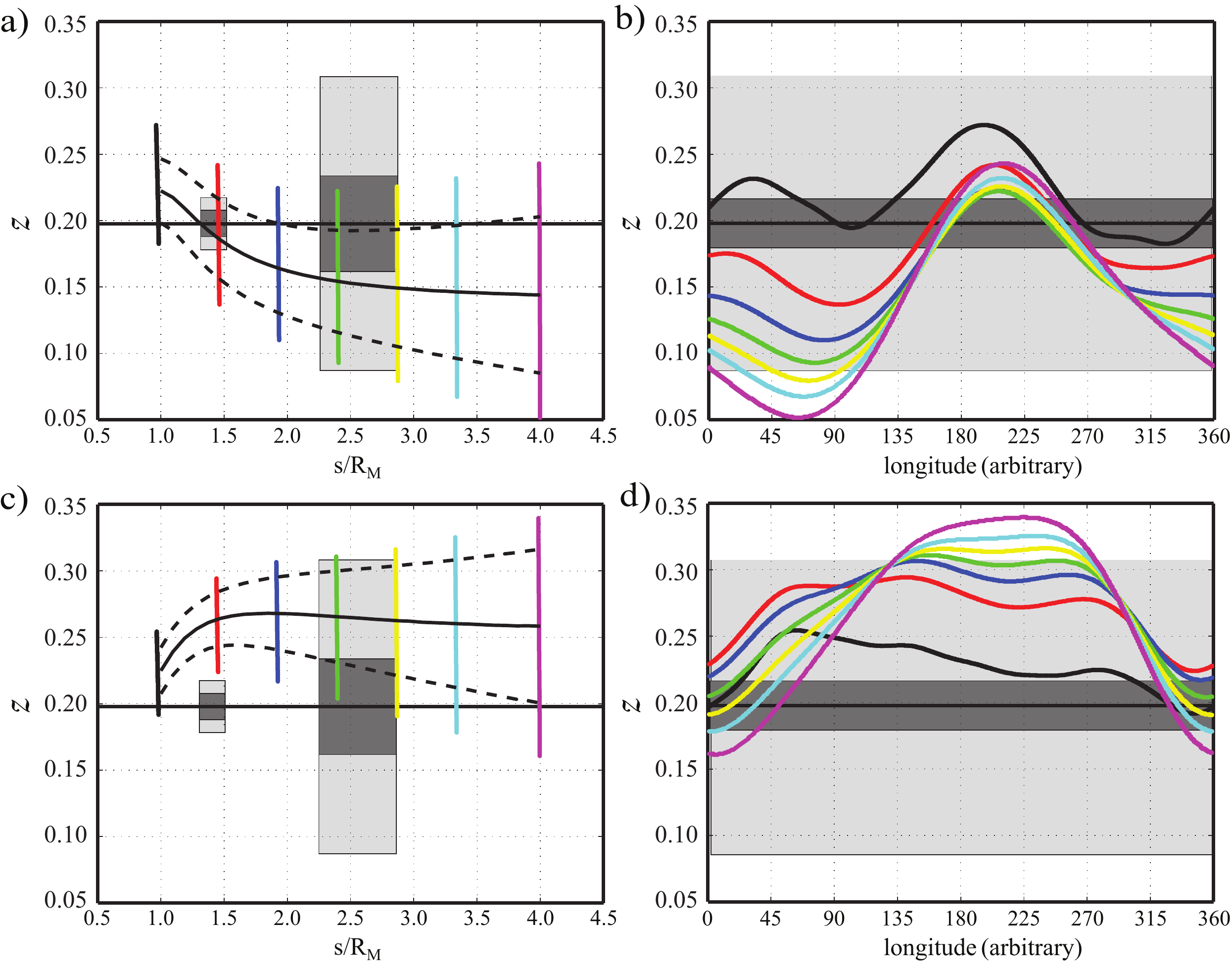}
\caption{Magnetic equator location for two snapshots in dynamo models
CW3 (top panels) and CW4 (bottom panels). The respective snapshot
times have been marked by the vertical dashed lines in \figref{MerT_Uli}.
Coloured dots show the equator locations found on a dense
spherical longitude/latitude grid.
The curved solid lines in panels a) and b) show the
mean equator offset for each spherical surface with radius $s/R_M$,
the dashed lines show the mean offset
plus and minus the standard deviation. Thick horizontal lines illustrate the
mean offset measured by the MESSENGER magnetometer while
mid gray and light grey boxed show mean three sigma error and
standard deviation for descending (left) and ascending orbits (right),
respectively.}
\label{MEall_Uli}
\end{figure*}

\begin{figure*}
\centering
\includegraphics[draft=false,width=0.7\textwidth]{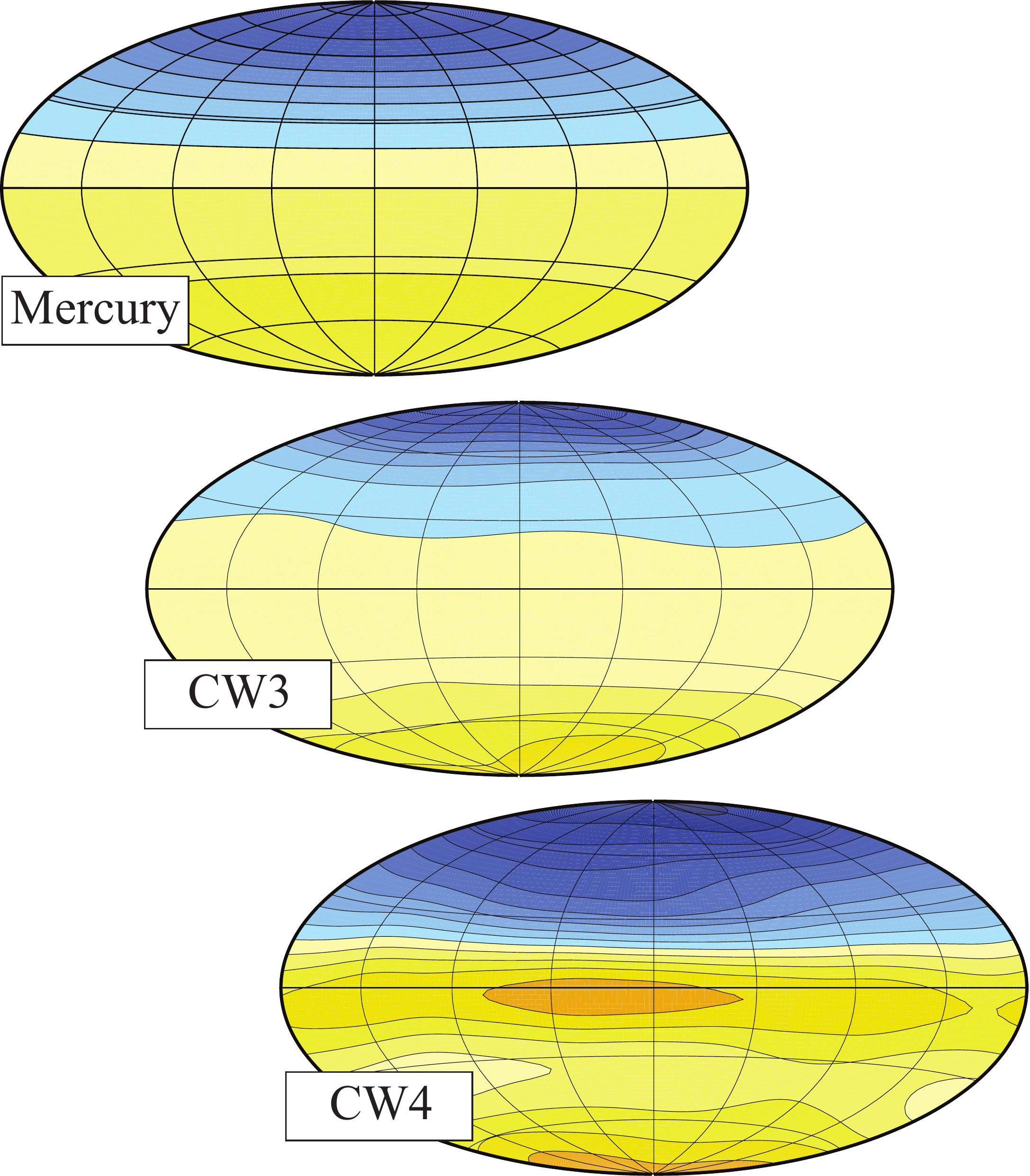}
\caption{Comparison of the MODM radial magnetic
field for Mercury with the two particularly Mercury-like
snapshots in models CW3 and CW4 already depicted in \figref{MEall_Uli}.
Blue (red and yellow) indicates radially
inward (outward) field.}
\label{Br_Sur_CW}
\end{figure*}

\Figref{MEall_Uli} illustrates the location of the magnetic equator for
two particularly Mercury-like snapshots in the two larger Rayleigh number
models CW3 and CW4.
\Figref{Br_Sur_CW} directly compares the respective radial magnetic fields
with the MESSENGER model. Both figures demonstrate that solutions very
similar to the offset dipole field proposed for Mercury can be found
with a stably stratified outer core layer and a sufficiently
high Rayleigh number.
However, the magnetic field varies considerably in time
and since neither hemisphere is preferred the offset can switch
from north to south and back.
The particular offset dipole configuration encountered by MESSENGER would
thus only be transient and representative for only a few percent of the
time at best.

\begin{figure*}
\centering
\includegraphics[draft=false,width=0.6\textwidth]{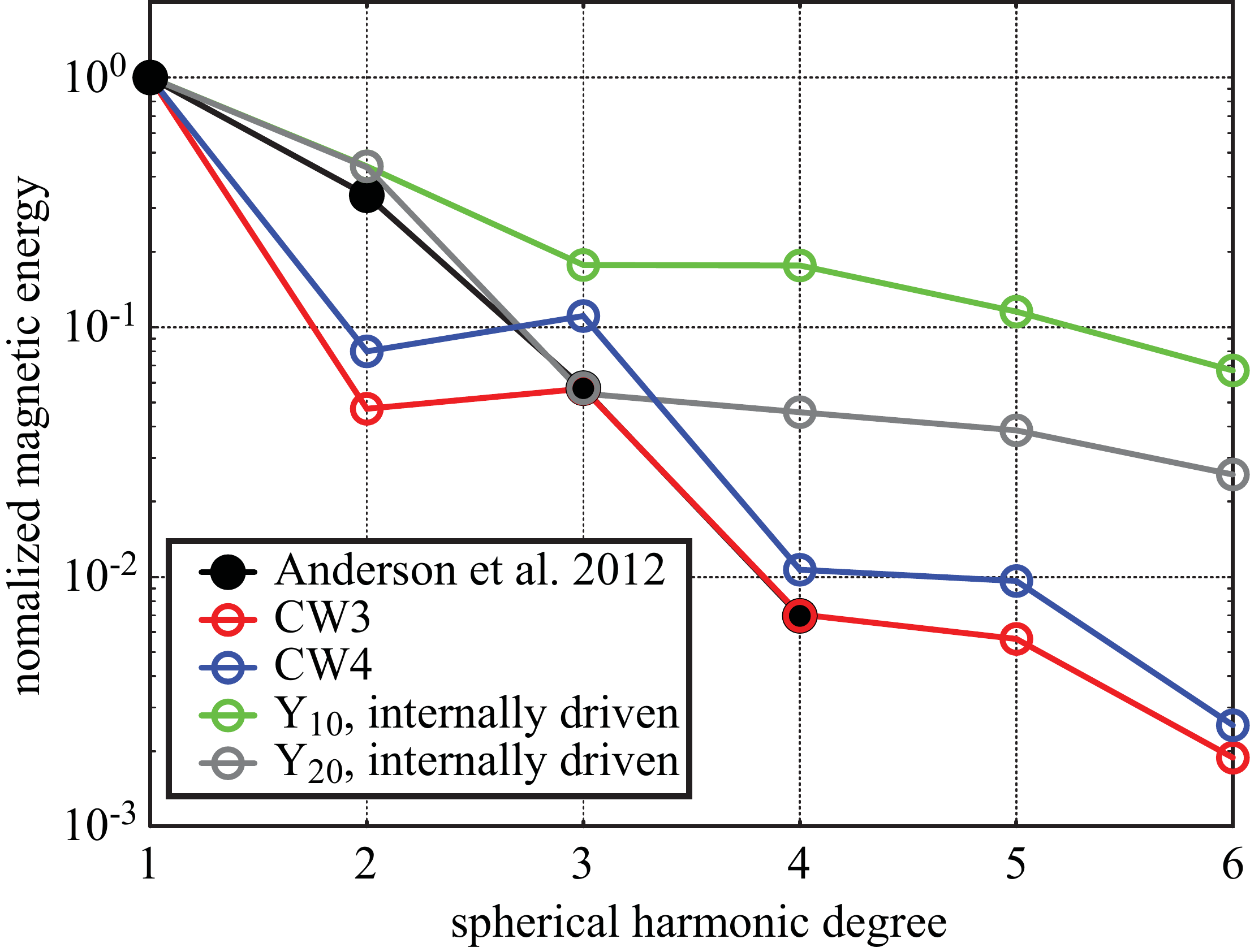}
\caption{Comparison of the normalized MODM surface spectrum by \citet{Anderson2012}
with time averaged spectra for four different dynamo models:
the dynamo models CW3 and CW4 that incorporate a stably stratified outer layer
\citep{Christensen2008} and models with an inhomogeneous core-mantle
boundary heat flux following a spherical harmonic $Y_{10}$ or $Y_{20}$
pattern, respectively.}
\label{Spectra1}
\end{figure*}

\Figref{Spectra1} compares the time averaged spherical harmonics surface
spectrum of models CW3 and CW4 with MODM, confirming that the relative
quadrupole contribution and thus the equator offset is typically too low.
The relative energy in spherical harmonic degrees $\ell=3$ and $4$, however,
agrees quite well with MESSENGER observations.

\citet{Manglik2010} explore what happens to the stable layer when giving
up the codensity formulation. They use a so-called
double-diffusive approach where two equations of the form of \eqnref{eqn:heat}
separately describe the evolution of temperature and composition.
When assuming a compositional diffusivity that is one order of magnitude lower
than the thermal diffusivity, the compositional plumes that rise from the inner core
boundary already stay significantly narrower than their thermal counterparts.
This allows them to more easily penetrate and destroy the stable outer layer.
The desirable filtering effect is greatly lost unless the sulphur
concentration is below $1\,$wt\% where compositional convection starts to
play an inferior role. For such a low light element concentration, however,
Mercury's core would likely be completely solid today.

The iron snow mechanism discussed in \secref{Interior}
offers an alternative scenario where the stable stratified layer is
likely to persist even in a double-diffusive approach.
The sulfur gradient that develops in the iron snow zone is potentially much
more stabilizing than the sub-adiabatic thermal gradient assumed
by \citet{Christensen2008} and \citet{Manglik2010}.
Furthermore, the additional convective driving source represented
by the remelting snow would counteract the effects of the more sulfur
rich plumes rising from a growing inner core.

\subsection{Inhomogeneous boundary conditions}

As already discussed in \secref{Intro}, an inhomogeneous heat flux through
the CMB is an obvious way to break the north/south symmetry and inforce
a more permanent offset of the magnetic equator.
To explain the stronger magnetization of the southern crust on Mars
several authors explored a variation following a
spherical harmonic function $Y_{10}$ of degree $\ell=1$
and order $m=0$ \citep{Stanley2008,Amit2011,Dietrich2013}.
The total CMB heat flux is then given by
\bel{HF}
q = q_0 \left( 1 - q_{10}^\star \cos{}\theta\right)
\ee
where $q_0$ is the mean heat flux, $q_{10}^\star$ the relative
amplitude of the lateral variation, and $\theta$ the colatitude. Positive values
of $q_{10}^\star$ are required at Mars and negative
values should enforce the stronger northern magnetic field
observed on Mercury.

\begin{figure*}
\centering
\includegraphics[draft=false,width=0.9\textwidth]{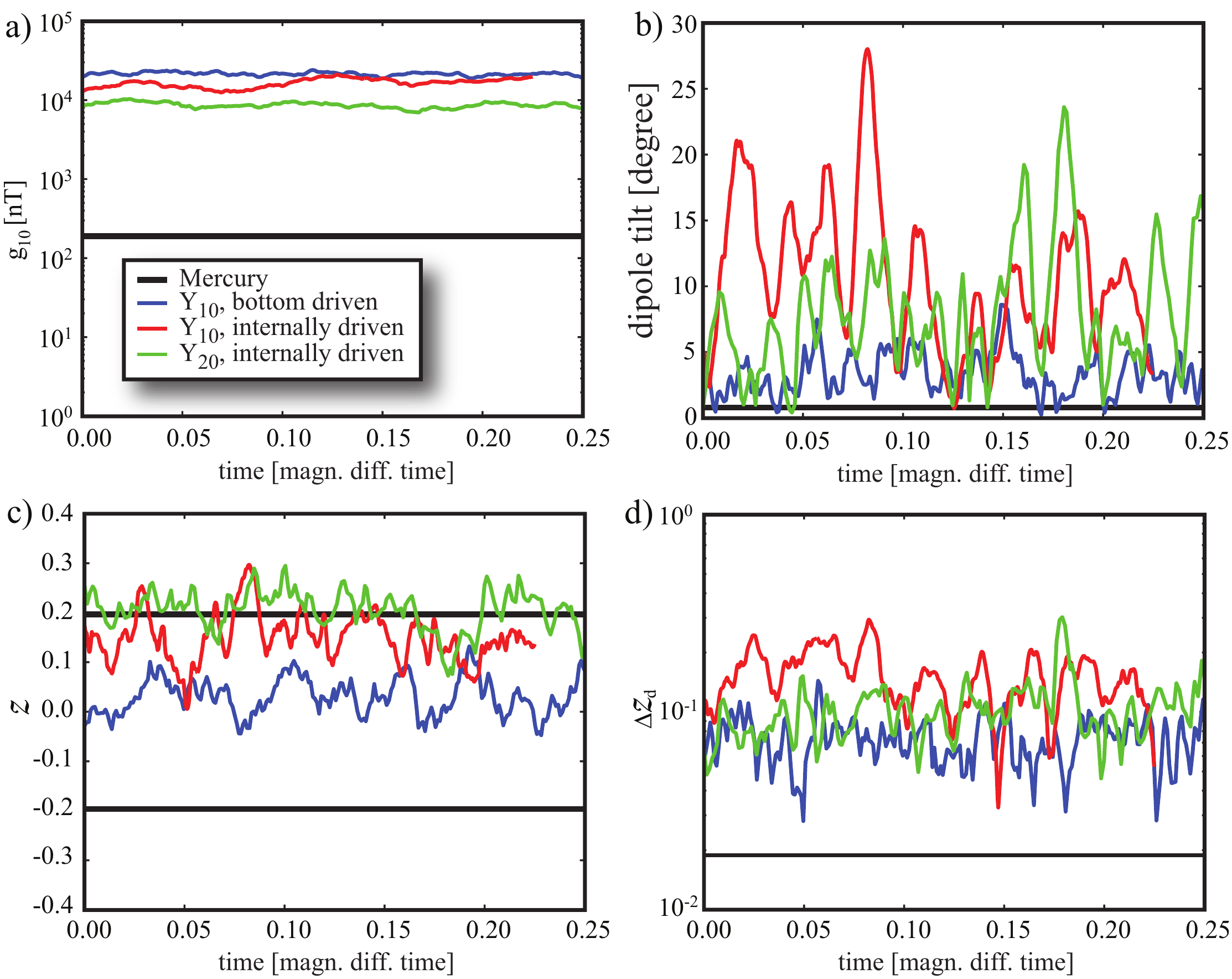}
\caption{Time evolution of three dynamo models with inhomogeneous
core-mantle boundary heat flux. A $Y_{10}$ pattern with
increased heat flux through the northern hemisphere but
also a $Y_{20}$ pattern with a larger heat flux in the equatorial
region promotes a Mercury-like offset.
See text and \figref{MerT_BH} for more explanation.}
\label{MerT_Ylm}
\end{figure*}

To explore the impact of the CMB heat flux pattern we use
dynamo simulations in the parameter range discussed by
\citet{Dietrich2013} and \citet{Cao2014}.
The parameters are $E=10^{-4}$, $\Ra=4\tp{7}$, $\Pra=1$, $\Pm=2$, and
$a=0.35$. Once more, rigid boundary conditions are used and we impose the heat
flux at the outer boundary. The Rayleigh number is then defined based on the
mean CMB heat flux \citep{Dietrich2013}.
\Figref{MerT_Ylm} demonstrates that a relative variation
amplitude of $q_{10}=-0.10$ is nearly sufficient to enforce the observed
offset when the dynamo is driven by homogeneously distributed
internal sources. These may either model secular cooling, radioactive
heating, or the remelting of iron snow. We have used a codensity formulation
here and set the codensity flux from the inner core boundary to zero.
Significantly larger heat flux variations are required for the
other end member when the dynamo is driven by bottom sources that mimic
a growing inner core.
This is consistent with the findings by \citep{Hori2012}
who report that the impact of thermal CMB boundary conditions is generally
larger for internally driven than bottom driven simulations.

Another not so obvious method to promote a north/south asymmetry is to
increase the heat flux through the equatorial region.
\citet{Cao2014} explore a $Y_{20}$ pattern, which means that the total CMB flux
is given by
\bel{HF2}
q = q_0 \left( 1 - q_{20}^\star
\frac{1}{2}\,(3\cos{}\theta-1)\,\right)\;\;.
\ee
The green line in \figref{MerT_Ylm} illustrates that a variation
amplitude of $q_{20}^\star=1/3$ causes a more or less
persistent Mercury-like offset value.
This translates into an increase of the equatorial flux by  $17$\% and
a decreases to polar flux by $33$\%. Cases with an increased heat
flux at the poles, i.e.~negative values of $q_{20}^\star$, did not
yield the desired result.
Except for the CMB heat flux pattern and a smaller inner core
that only occupies 20\% of the radius, the models are identical
to the $Y_{10}$ cases explored above.
Though the $Y_{20}$ pattern is equatorially symmetric, it promotes an
equatorially asymmetric flow and therefore an asymmetric magnetic
field production. A preliminary analysis of the system suggest that the
$Y_{20}$ pattern significantly decreases the critical Rayleigh number
for the onset of equatorially anti-symmetric convection modes,
which is very large when the CMB heat flux is homogenous \citep{Landeau2011}.

The inhomogenous CMB heat flux mainly helps to promote a Mercury-like
mean offset of the magnetic equator while other important
field characteristics seem not consistent with the observations.
The field is generally much too strong and the often large dipole tilt and
offset spread $\Delta\mathcal{\overline{Z}}$ testify
that higher harmonic and non-axisymmetric field contributions
remain too significant. This is confirmed by the time averaged
spectra shown in \figref{Spectra1}.
Adding a stably stratified outer layer, probably in combination
with a larger Rayleigh number to bring down the too stong axial dipole
contribution, seems like an obvious solution to this problem.
This was confirmed by the first results presented by \citet{Tian2013}
who explore the combiation of the stable layer with the
$Y_{10}$ heat flux pattern.

\subsection{Alternatives}

\begin{figure*}
\centering
\includegraphics[draft=false,width=8cm]{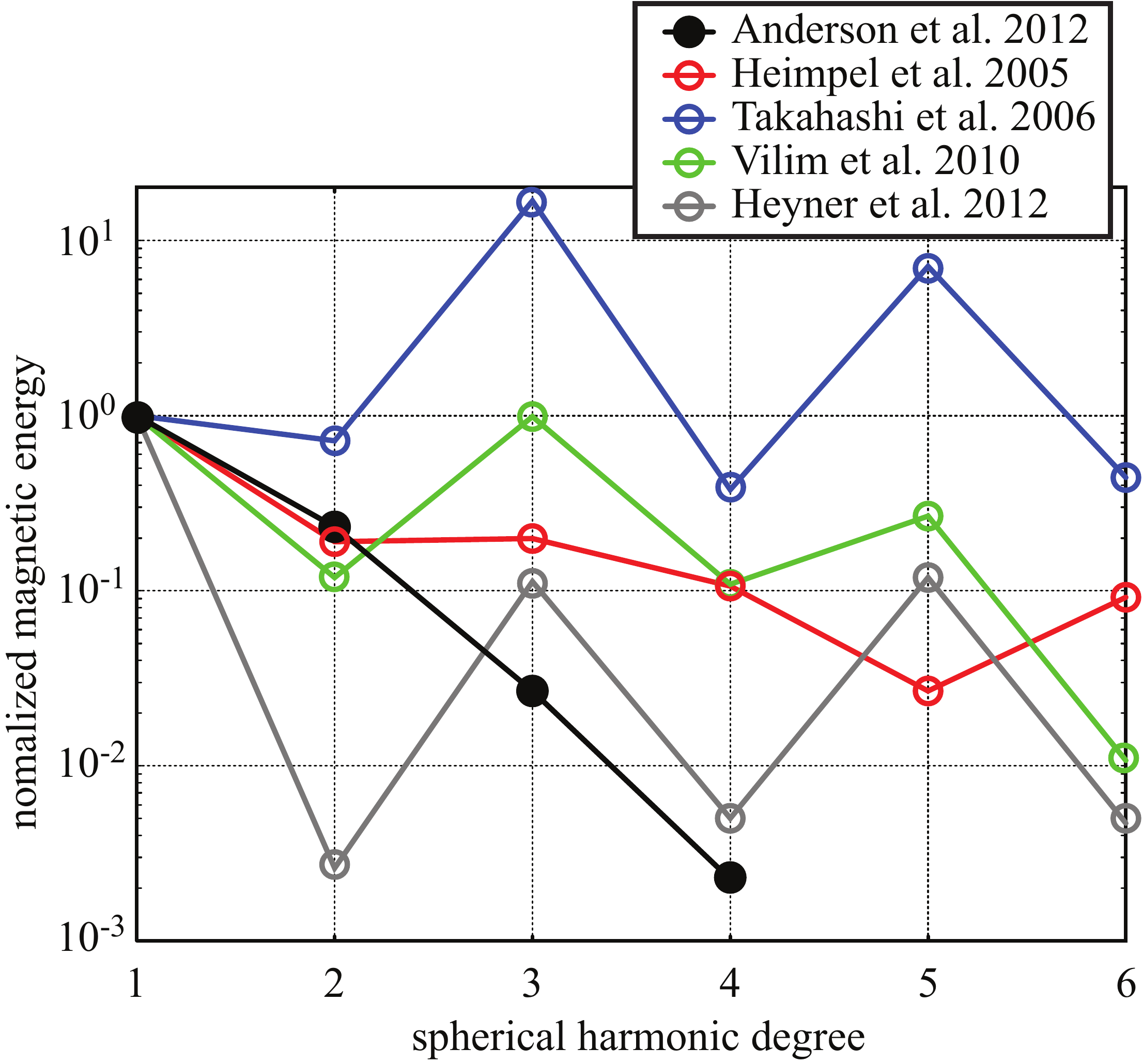}
\caption{Comparison of the normalized MODM surface spectrum by \citet{Anderson2012}
with spectra for different dynamo models.
A time averaged spectrum is shown for the models by \citet{Vilim2010}
and \citet{Heyner2011} while
the spectra for \citet{Heimpel2005a}, \citet{Takahashi2006}
represent snapshots.}
\label{Spectra2}
\end{figure*}

Several authors varied the inner core size to explore its impact
on the dynamo process. \citet{Heimpel2005a} analyse models with aspect ratios
between $a=0.65$ and $a=0.15$ that are all driven by a growing inner core.
They report that the smallest inner core yields a particularly weak
magnetic field with a CMB Elsasser number of $\Lambda_{cmb}=10^{-2}$
when the Rayleigh number is close to onset for dynamo action.
This is still more than two orders of magnitude too large for Mercury.
Convection and dynamo action are mainly concentrated at only one convective
column attached to the inner core.
Such localized magnetic field production is not very conducive to
maintaining a large scale magnetic field which is confirmed
by the magnetic surface spectrum of a model snap shot shown in \figref{Spectra2}.
The relative quadrupole contribution nearly matches the MODM
value but the higher harmonic contributions can reach a similar
level and are thus too strong. This is also true for the dipole tilt
which has a mean value of $8^\circ$ for this model.

\citet{Takahashi2006} find that the magnetic field strength
is also reduced when using a large inner core
with $a=0.7$ in combination with a large Rayleigh number.
Once more, the field is still too strong for Mercury
with an Elsasser number around $\Lambda_{cmb}=10^{-2}$ and
is also much too small in scale
 with $\ell=3$ and $5$ contributions dominating
 the spectrum (see \figref{Spectra2}).
\citet{Stanley2005} explore even larger inner cores
with aspect ratios up to $a=0.9$ and report particularly
weak fields at rather low Rayleigh numbers.
The use of stress-free flow boundary conditions set this dynamo
model apart from all the other cases discussed here.
Field strength, dipole tilt, and offset are highly variable
but seem to assume Mercury-like values at times.
Little more is published about the field geometry and it
seems worth to explore these models further.

\citet{Vilim2010} explore the double snow zone regime that
may develop when the sulfur content in Mercury's core
exceeds $10\,$wt\%, as briefly discussed in \secref{Internal}.
They consider a thin outer snow zone and a thicker zone in the middle
of the liquid core in addition to a growing inner core.
Since both snow zones are stably stratified the
dynamo action is concentrated in the two remaining shells.
The magnetic fields that are produced in these two dynamo regions tend
to oppose each other which leads to a reduced overall field strength that
matches the MESSENGER observation.
However, the octupole component is generally too strong
while the quadrupole contribution is too weak,
as is demonstrated in \figref{Spectra2}.

Since internal and external magnetic field can reach similar
magnitudes at Mercury the latter may actually play a role
in the core dynamo. The idea of a feedback between
internal and external dynamo processes was first proposed
by \citet{Glassmeier2007} for Mercury and further developed in a
series of papers \citep{Heyner2010,Heyner2011,Heyner2011a}.
Because the internal dynamo process operates on time scales
of decades to centuries, only the long time-averaged
magnetospheric field needs to be considered.
This can be approximated by an external axial dipole that opposes
the direction of the inernal axial dipole within the core.
The ratio of the external to internal dipole field
depends on the distance of the magnetopause to the planet and
thus on the intensity of the internal field.
\citet{Heyner2011} find that the feedback quenches the dynamo field
to Mercury-like intensities when the simulation is started off with
an already weak field and the Rayleigh number is not too high.
These conditions can, for example, be met when dynamo action is
initiated with the beginning of iron snow or inner core growth
at a period in the panetary evolution where mantle convection
is already sluggish and the CMB heat flux therefore low.
The feedback process modifies the dipole dominated
field by concentrating the flux at higher latitudes. The result
is a spectrum where the relative quadrupole
(and other equatorially anti-symmetric contributions) is too weak while
the octupole (and other equatorially symmetric contributions) is too
strong (see \figref{Spectra2}).

\section{Conclusion}
\label{Conclusion}

The MESSENGER data have shown that Mercury has an exceptional magnetic
field \citep{Anderson2012,Johnson2012}.
The internal field is very weak and has a simple but
surprising geometry that is consistent with an axial dipole
offset by $20\%$ of the planetary radius to the North.
This implies a very strong axial quadrupole but at the same time also
small higher harmonic and non-axial contributions, a unique
combination in our solar system.

Numerical dynamo models have a hard time to explain these observations.
Strong axial quadrupole contributions and thus a significant mean
offset of the magnetic equator can be promoted by different
measures. Very small and very large inner cores or strong inertial forces
are three possibilities that lead to a sizable but also very
time dependent axial quadrupole contribution.

A more persistent Mercury-like mean offset can be enforced by
imposing lateral variations in the core-mantle boundary heat flux.
Pattern with either an increased heat flux in the northern hemisphere
or in the equatorial region yield the desired result.
They are particularly effective when the dynamo is not driven
by a growing inner core but by homogeneously distributed buoyancy sources
\citep{Cao2014}.
New models for Mercury's interior, however, suggest that neither
pattern is likely to persist today.

Unfortunately, the measures that promote a stronger axial quadrupole also
tend to promote non-dipolar and non-axisymmetric field contributions in
general. The offset of the magnetic equator therefore
strongly depends on longitude and distance to the planet, which is
at odds with the MESSENGER observations.
Dynamo simulations by \citet{Christensen2006b} and \citet{Christensen2008}
have shown that a stably stratified outer core layer helps to solve
this problem. The magnetic field that is produced in the deeper
core regions has to diffusive through this largely passive
layer to reach the planetary surface. And since the magnetic field
varies in time it is damped by the magnetic skin effect during this
process. Higher harmonic and non-axisymmetric contributions
are damped more effectively than axial dipole or quadrupole because
the variation time scale decreases with increasing spatial complexity.
When reaching the surface, the field is therefore not only more Mercury-like
in geometry but also similarly weak.

Recent interior models for Mercury suggest that a stable outer core layer
may indeed exist. Because of the low pressures in
Mercury's outer core, an outer iron snow zone should develop
underneath the CMB for mean core sulfur concentration beyond about $2\,$wt\%.
As the planet cools, the snow zone extends deeper into the
core and a stably stratifying sulfur gradient develops.
Since the mean heat flux out of the Hermean core is likely subadiabatic
today, thermal effects would further contribute to stabilizing
the outer core region. Such a layer is also likely to
persist when double-diffusive effects are taken into account
\citep{Manglik2010}.
Additional work on the FeS melting behaviour, on Mercury's interior
properties, and the planet's thermal evolution is required to
better understand and establish this scenario.
The possible presence of Si in the Hermean core could further
complicate matters \citep{Malavergne2010}.

Dynamo simulations that more realistically model the iron snow
stratification and the convective driving in the presence
of an iron snow zone and possibly also a growing inner core seem
a logical next step. Lateral variations in the core-mantle
boundary heat flux and a feedback with the magnetospheric field
are two other features that may play an important role
in Mercury's dynamo process.

The Hermean magnetospheric field remains a challenging puzzle despite the
wealth of data delivered by the MESSENGER magnetometer. It's small size
and high variability complicates the separation of internal and
external field contributions, of temporal and spatial variations,
and of solar wind dynamics and Mercury's genuine field dynamics.
The BepiColombo mission, scheduled for launch in 2016,
will significantly improve the situation since two
spacecrafts will orbit the planet at the same time,
a planetary orbiter build by ESA and a magnetospheric orbiter build by JAXA.


\section*{Acknowledgement}
Johannes Wicht was supported by the Helmholz
Alliance "Planetary Evolution and Live" and by the Special Priority
Programm 1488 "Planetary Magnetism" of the German Science Foundation.
D. Heyner was supported by the German Ministerium f\"ur
Wirtschaft und Technologie and the German Zentrum f\"ur Luft- und Raumfahrt
under contract 50 QW 1101.
We thank Attilio Rivoldini, Tina R\"uckriehmen, Wieland Dietrich, Hao Cao,
Brian Anderson, Karl-Heinz Glassmeier, and Ulrich R.~Christensen
for helpful discussions. Attilio Rivoldini also kindly provided figure 1.

\bibliographystyle{unsrtnat}
\bibliography{JWbib}

\end{document}